\documentclass[11pt]{amsart}
\usepackage{graphicx,amssymb,amsmath,amsthm}
\usepackage{enumerate}
\usepackage{dsfont}
\usepackage[colorlinks, citecolor=red]{hyperref}
\usepackage{comment,cite,color}
\usepackage{cite,color}
\usepackage{mathrsfs}
\usepackage{epsfig}
\usepackage{lscape}
\usepackage{subfigure}
\usepackage{epstopdf}
\usepackage{caption}
\usepackage{bm}
\usepackage{algorithm}
\usepackage{algpseudocode}
\usepackage[english]{babel}
\usepackage{multirow}

\newcommand{\xkh}[1]{\left(#1\right)}

\newcommand{\zkh}[1]{\left[#1\right]}

\newcommand{\nj}[1]{\langle {#1} \rangle}

\newcommand{\norm}[1]{\|{#1}\|}

\newcommand{\norms}[1]{\|{#1}\|}
\newcommand{\abs}[1]{\lvert#1\rvert}
\newcommand{\Abs}[1]{\left\lvert#1\right\rvert}

\newcommand{\ra}{{\rightarrow}}
\newcommand{\lra}{{\longrightarrow}}
\newcommand{\ep}{{\varepsilon}}
\newcommand{\eproof}{\hfill\rule{2.2mm}{3.0mm}}

\newcommand{\Proof}{\noindent {\bf Proof.~~}}

\newcommand{\E}{{\mathbb E}}

\newcommand{\1}{{\mathds 1}}

\newcommand{\R}{{\mathbb R}}
\newcommand{\Rn}{{\mathbb R}^n}

\newcommand{\T}{\top}

\newcommand{\vx}{{\bm x}}

\newcommand{\vv}{{\bm v}}
\newcommand{\vw}{{\bm w}}
\newcommand{\vz}{{\bm z}}

\newcommand{\va}{{\bm a}}

\newcommand{\sgn}{{\rm sgn}}

\newcommand{\supp}{{\rm supp}}

\newcommand{\RNum}[1]{\uppercase\expandafter{\romannumeral #1\relax}}

\newtheorem{definition}{Definition}[section]
\newtheorem{corollary}[definition]{Corollary}
\newtheorem{prop}[definition]{Proposition}
\newtheorem{theorem}[definition]{Theorem}
\newtheorem{lemma}[definition]{Lemma}

\newtheorem{remark}[definition]{Remark}

\newtheorem{example}[definition]{Example}

\date{}

\begin{document}

\author{Jianfeng Cai}
\thanks{J. F. Cai was supported in part by Hong Kong Research Grant Council grants 16306317 and 16309219}
\address{Department of Mathematics, The Hong Kong University of Science and Technology, Clear Water Bay, Kowloon, Hong Kong}
\email{jfcai@ust.hk}

\author{Meng Huang}
\address{Department of Mathematics, The Hong Kong University of Science and Technology, Clear Water Bay, Kowloon, Hong Kong}
\email{menghuang@ust.hk}

\author{Dong Li}
\thanks{D. Li was supported in part by Hong Kong RGC grant GRF 16307317
 and 16309518.}
\address{Department of Mathematics, The Hong Kong University of Science and Technology, Clear Water Bay, Kowloon, Hong Kong.}
\email{madli@ust.hk}

\author{Yang Wang}
\thanks{Y. Wang was supported in part by the Hong Kong Research Grant Council grants 16306415 and 16308518.}
\address{Department of Mathematics, The Hong Kong University of Science and Technology, Clear Water Bay, Kowloon, Hong Kong}
\email{yangwang@ust.hk}

\baselineskip 18pt
\bibliographystyle{plain}
\title[Solving phase retrieval with random initial guess]{Solving phase retrieval with random initial guess  is nearly as good as by spectral initialization }\footnote{The results in our paper were first presented at a workshop in December, 2019. }
\maketitle

\begin{abstract}
The problem of recovering a signal $\vx\in \Rn$ from a set of magnitude measurements $y_i=\abs{\nj{\va_i,\vx}}, \; i=1,\ldots,m$  is referred as phase retrieval, which has many applications in fields of physical sciences and engineering.
In this paper we show that the smoothed  amplitude flow model for phase retrieval has benign geometric structure under the optimal sampling complexity. In particular, we show that when the measurements $\va_i\in \Rn$ are Gaussian random vectors and the number of measurements $m\ge Cn$, our  smoothed  amplitude flow model has no spurious local minimizers with high probability, ie., the target solution $\vx$ is the unique global minimizer (up to a global phase) and the loss function has a negative directional curvature around each saddle point.  Due to this benign geometric landscape, the phase retrieval problem can be solved by  the gradient descent algorithms without spectral initialization. Numerical experiments show that the gradient descent algorithm with random initialization performs well even comparing with state-of-the-art algorithms with spectral initialization in empirical success rate and convergence speed.
\end{abstract}

\section{Introduction}
\subsection{Background}
This paper concerns the well-known phase retrieval problem, which aims to recover the signal $\vx\in \Rn$ from a series of magnitude-only measurements 
\[
y_i=\abs{\nj{\va_i,\vx}}, ~i=1,\ldots,m
\]
where $\va_i\in \Rn, i=1,\ldots,m$ are Gaussian random vectors and $m$ is the number of measurements. This problem arises in many fields of science and engineering due to the physical limitations of optical detectors which can only record the magnitude of signals while losing the phase information, such as X-ray crystallography \cite{harrison1993phase,millane1990phase}, microscopy
\cite{miao2008extending}, astronomy \cite{fienup1987phase}, coherent diffractive
imaging \cite{shechtman2015phase,gerchberg1972practical} and optics
\cite{walther1963question} etc. Despite its simple mathematical form, it has been
shown that reconstructing a finite-dimensional discrete signal from the magnitude of its Fourier transform is generally an {\em NP-complete} problem \cite{Sahinoglou}.

Many algorithms have been designed to solve the phase retrieval problem. They fall generally into two categories: convex algorithms and non-convex ones.  The convex algorithms usually rely on a ``matrix-lifting'' technique, which lifts the phase retrieval problem into a low rank matrix recovery problem, together with convex relaxation by showing that the matrix recovery problem under some conditions is equivalent to a convex optimization problem. These algorithms include PhaseLift \cite{phaselift,Phaseliftn}, PhaseCut \cite{Waldspurger2015} etc. It has been shown \cite{Phaseliftn} that PhaseLift can achieve the exact recovery under the optimal sampling complexity with Gaussian random measurements.

Although convex methods have good theoretical guarantees to converge to the true solutions under some special conditions, they tend to be computationally inefficient for large scale problems. By contrast, many non-convex algorithms do not need the lifting step so they operate directly on the lower-dimensional ambient space, making them much more efficient. Early non-convex algorithms were based mostly on alternating projections, e.g. Gerchberg-Saxton \cite{Gerchberg1972} and Fineup \cite{ER3}. The drawback is the lack of theoretical guarantee. Later Netrapalli et al \cite{AltMin} proposed the AltMinPhase  algorithm based on a technique known as {\em spectral initialization}, and they proved that the algorithm linearly converges to the true solution with $O(n \log^3 n)$ resampling Gaussian random measurements. This work led to further several other non-convex algorithms based on spectral initialization. They share the common idea of first choosing a good initial guess through spectral initialization, and then solving an optimization model through gradient descent. Two commonly used optimization models are the intensity flow model
\begin{equation}\label{eq:mod1}
\min_{\vz\in \R^d}\,\,F(\vz)=\frac{1}{m}\sum_{j=1}^m \xkh{\abs{\nj{\va_j,\vz}}^2-y_j^2}^2.
\end{equation}
and the amplitude flow model
\begin{equation}\label{eq:mod2}
\min_{\vz\in \R^d}\,\,F(\vz)=\frac{1}{m}\sum_{j=1}^m \xkh{\abs{\nj{\va_j,\vz}}-y_j}^2,
\end{equation}
Specifically,  Cand\`es et al developed the Wirginger Flow (WF)  \cite{WF} method based on  (\ref{eq:mod1}) and proved that the WF algorithm can achieve linear convergence with $O(n \log n)$ Gaussian random measurements.  Lately, Chen and Cand\`es  improved the results to $O(n)$ Gaussian random measurements by incorporating a truncation, namely  the Truncated Wirginger Flow (TWF) \cite{TWF} algorithm. Other methods based on (\ref{eq:mod1}) include the Gauss-Newton \cite{Gaoxu} method, the trust-region \cite{turstregion} method, and others.  Several algorithms based on the amplitude flow model (\ref{eq:mod2}) have also been developed recently, such as the Truncated Amplitude Flow (TAF) algorithm \cite{TAF}, the Reshaped Wirtinger Flow (RWF) \cite{RWF} algorithm and the Perturbed Amplitude Flow (PAF) \cite{PAF} algorithm. All three algorithms above have been shown to linearly converge to the true solution up to a global phase with $O(n)$ Gaussian random measurements. Furthermore, numerical results show that algorithms based on the amplitude flow model (\ref{eq:mod2}) tend to outperform algorithms based on model (\ref{eq:mod1}).

\subsection{Motivation and Related Work}
As we have stated earlier, producing a good initial guess using spectral initialization is a prerequisite for all aforementioned non-convex algorithms with theoretical guarantee. An interesting question is that {\em Is it possible for those algorithms to achieve successful recovery with a random initialization}?  


For intensity-based model (\ref{eq:mod1}), the answer is yes.  Ju Sun et al \cite{turstregion} study the global geometry structure of the loss function of (\ref{eq:mod1}). They show the loss function $F(\vz)$  does not have any spurious local minima under $O(n \log^3 n)$ Gaussian random measurements. It means that all
minimizers are the target signal $\vx$ up to a global phase and the loss function has a negative directional curvature around each saddle point. Thus any algorithm which can avoid saddle points  converges to the true solution with high probability. They also develop a trust-region 
method to find a global solution with random initialization. To reduce the sampling complexity, it has been shown that the combination of the loss function (\ref{eq:mod1}) and an activation function also possesses the benign geometry structure  under $O(n)$ Gaussian random measurements \cite{cai2019}.

The geometry landscape concept has also been explored in recent years for other applications in signal processing and machine learning, e.g. matrix sensing \cite{bhojanapalli2016global,park2016non}, tensor decomposition \cite{ge2016matrix}, dictionary learning\cite{sun2016complete} and matrix completion \cite{ge2015escaping}. Well-behaved geometry landscapes for optimization, namely all local optimal are also global optimal and the loss function has a negative directional curvature around each saddle point, have been shown to exist more broadly. Several techniques have been developed to guarantee that the basic gradient optimization algorithms can escape such saddle points efficiently, see e.g. \cite{jin2017escape,du2017gradient,jin2017accelerated}.

\subsection{Our contributions}

Optimization algorithms based on the amplitude model (\ref{eq:mod2}) have been shown to outperform those based on the intensity model (\ref{eq:mod1}). Naturally we may ask whether it is possible to examine the geometric landscape for  the amplitude model (\ref{eq:mod2})  and develop algorithms similar to the ones in \cite{turstregion, cai2019}. As it turns out, a straightforward approach based on  model (\ref{eq:mod2}) loss function fails as there will be many local minima regardless how many measurements one take.  In this paper, we show that by altering the amplitude model based loss function slightly we are able to obtain to benign geometric landscape for the loss function, thus yielding a fast algorithm that requires only $O(d)$ measurements and no initialization. Furthermore, numerical tests show that the algorithm outperforms several existing algorithms in terms of efficiency.

We now describe our study in more details. Let $\vx \in \Rn$ be the target signal we want to recover. The measurements we obtain are 
\[
     y_i=\abs{\nj{\va_i,\vx}}, ~i=1,\ldots,m
\]
where $\va_i\in \Rn, i=1,\ldots,m$ are Gaussian random vectors. For the recovery of $\vx$ we consider the following new  loss function $F(\vz)$ given by
 \begin{equation} \label{mo:mymodel}
F(\vz)=\frac{1}{2m}\sum_{i=1}^m \xkh{\gamma\xkh{\frac{\abs{\va_i^\T \vz}}{\abs{\va_i^\T \vx}}}-1}^2\cdot \abs{\va_i^\T \vx}^2,
\end{equation}
where the function $\gamma(t)$ is taken to be 
\begin{equation}\label{eq:gamma}
\gamma(t):=\left \{ \begin{array}{cl}
     \abs{t}, &  \abs{t} > \beta;  \\
    \frac{1}{2\beta} t^2+\frac{\beta}{2}, &  \abs{t} \leq \beta .
\end{array} \right. 
\end{equation}
Note that the event $\bigcup_{i=1}^m\{ \va_i^\T \vx =0 \}$ has zero probability and we may
assume that $\va_i^T \vx \neq 0$ for all $i$. 
Another practical way is to define $(\gamma(\frac{\abs{\va_i^\T \vz}}{\abs{\va_i^\T \vx}})-1)^2\cdot \abs{\va_i^\T \vx}^2=  \abs{\va_i^\T \vz}^2$
when $\va_i^\T \vx=0$.

Because $\gamma(t)$ is smoothed from $|t|$ in the amplitude flow model, we shall call our model the  {\em Smoothed  Amplitude Flow (SAF)} model.  Clearly if we set $\gamma(t) :=|t|$ then the loss function is exactly the amplitude flow model loss function given in (\ref{eq:mod2}). Unfortunately in this case the loss function does not yield the desired  geometric landscape: Regardless how many measurements one take, there will appear multiple local minima. Our new loss function, however, will have the desired property. The main theorem of the paper is the following:

\begin{theorem} \label{theo-1.1}
Fix $0<\beta\le \frac 12$. 
  Assume $m\ge C n$. Let $\vx\in\Rn$ be nonzero and $\{\va_i\}_{i=1}^m$ be i.i.d. random Gaussian vectors, i.e.,  $\va_i \sim N(0, I_n)$ for all $i$. Then with probability at least $1-c\exp(-\delta m)$ the loss function $F(\vz)$ given in (\ref{mo:mymodel}) has no spurious local minima, i.e. all local minima are also global minima and any other critical point is a saddle point with a negative directional curvature. Here $C,c,\delta$ are  positive constants depending on $\beta$. 
\end{theorem}

\begin{remark}
For simplicity, we only consider the geometric landscape in the real case, however, the result in Theorem \ref{theo-1.1} can be adapted to the complex case and we will address it elsewhere.
\end{remark}
Theorem \ref{theo-1.1} implies that gradient descent with any random initial point will not get stuck in a local minimum. Our result turns out to be not just of theoretical interest. Numerical tests show that this model yields very stable and fast convergence with random initialization with performance on a par with or even better than the existing gradient descent methods with spectral initialization.

\subsection{Organization}
The rest of this paper is organized as follows. In Section 2, we provide an outline of the proof.  In Section 3, we break down $\Rn$ into several regions and investigate the geometric property of $F(\vz)$ on each region.  In Section 4, we carry out some numerical experiments to demonstrate the effectiveness of our model.  The appendix
collects the proofs of some technical lemmas and propositions.


\section{Geometric Properties of the SAF Loss Function} \label{geometry}

Our main theorem is a consequence of the analysis of the geometric landscape of the smoothed amplitude flow model loss function $F(\vz)$ in (\ref{mo:mymodel}). As with \cite{turstregion}, we shall decompose $\Rn$ into several regions (not necessarily non-overlapping), on each of which $F(\vz)$  has certain property that will allow us to show that with high probability $F(\vz)$ has no  local minimizers other than $\pm \vx$. Furthermore, we show $F(\vz)$ is strongly convex in a neighborhood of $\pm\vx$. 

Thus our strategy for proving the main result is as follows:
\begin{description}
\item[Step 1]  Compute the gradient and Hessian of the loss function $F(\vz)$. Since $F(\vz)$ is not 2nd order differentiable we shall consider the directional second derivative of $F(\vz)$. Notice that all these are given by  sums of random variables. 
\item[Step 2]  Apply concentration inequalities such as Bernstein's inequality as well as union bounds to approximate the sums of random variables for a given $\vz$. 
\item[Step 3]   Estimate the approximations obtained from concentration inequalities to establish the geometric properties for $F(\vz)$. In particular we shall estimate $\nj{\nabla F(\vz), \vz}$, $\nj{\nabla F(\vz), \vx}$, and $D^2_\vv F(\vz)$ where $D^2_\vv$ denotes the directional 2nd order derivative along the direction $\vv$ of $F(\vz)$.
\end{description}
Because $\gamma(t)$ of (\ref{eq:gamma}) is given piecewise, the main difficulty here lies with estimations in step 3. Fortunately, while tedious, they can be done to yield what we will need to prove the theorem.

\subsection{Step 1} Note that $F(\vz)$ is continuously differentiable, but it is not 2nd order differentiable, so for the 2nd order derivatives we will resort to directional derivatives. Recall that for a function $g(\vz)$ and any vector $\vv\neq 0$ in $\Rn$, the {\em one-side directional derivative of $g$} at $\vz$ along the direction $\vv$ is given by
\[
       D_{\vv} g(\vz):=\lim_{t\to 0^+} \frac{g(\vz+t\vv)-g(\vz)}{t}
\]
if the limit exists. Furthermore, we denote 
\[
       D_{\vv}^2 g(\vz)=D_{\vv}(D_{\vv} g(\vz))
\]
as the {\em second order directional derivative of $g$} at $\vz$ along the direction $\vv$. As we shall show, both the gradient $\nabla F(\vz)$ and the $D_\vv^2F(\vz)$ are subexponential random variables in terms of $\va^\T\vz$, $\va^\T\vx$, and $\va^\T\vv$. This enables us to apply concentration inequalities described in Step 2. The details will be given in Section 4.

\subsection{Step 2} Concentration inequalities allow us to estimate the sums of random variables so we can estimate them for proving the main result. We shall be dealing with subexponential random variables in this paper. 
A good discussion of subexponential random variables can be found in \cite{Vershynin2018}. 

\begin{lemma}[\cite{Vershynin2018}, Theorem 2.8.1]   \label{lem:concentration}
     Let $g(s,t)$ be a real valued function such that $g(\va_i^\T \vz, \va_i^\T \vx)$ is subexponential with subexponential norm $\|g(\va_i^\T \vz, \va_i^\T \vx)\|_{\Psi_1} \leq \tau$. Then for any $\varepsilon>0$ we have
\begin{equation}  \label{eq:concentration}
      \Bigl|\frac{1}{m} \sum_{i=1}^m  g(\va_i^\T \vz, \va_i^\T \vx) - \E[g(\va_1^\T \vz, \va_1^\T \vx)] \Bigr| \leq \varepsilon
\end{equation}
with probability at least  $1-2\exp(-cm\min(\varepsilon^2/\tau^2, \varepsilon/\tau))$, where $c>0$ is a universal constant.
\end{lemma}

Of course we will need  (\ref{eq:concentration}) to hold uniformly for {\em all} $\vz$ on certain region. This is typically proved by using some $\delta$-nets, and there is no exception here. We have

\begin{corollary}\label{coro:union}
     Let $g(s,t)$ be a real valued function such that $g(\va_i^\T \vz, \va_i^\T \vx)$ is subexponential with subexponential norm $\|g(\va_i^\T \vz, \va_i^\T \vx)\|_{\Psi_1} \leq \tau$. Assuming that on a compact set $\Omega$ we have
\begin{equation} \label{eq:lipschitz}
      \frac{1}{m} \sum_{i=1}^m \left| g(\va_i^\T \vz, \va_i^\T \vx)-g(\va_i^\T \vz_0, \va_i^\T \vx)\right| \leq \frac{1}{m} \sum_{i=1}^m K(\va_i) \abs{ \va_i^\T (\vz-\vz_0)}
\end{equation}
for any $\vz,\vz_0\in\Omega$, where $K(\va_i)$ is a subgaussian random variable with subgaussian norm $\eta$. 
Then for any $\varepsilon>0$ there exist constants $C,c>0$ depending only on $\varepsilon, \tau, \eta$ such that for $m \geq Cn$ we have
\begin{equation}  \label{eq:union}
      \Bigl|\frac{1}{m} \sum_{i=1}^m  g(\va_i^\T \vz, \va_i^\T \vx) - \E [g(\va_1^\T \vz, \va_1^\T \vx)] \Bigr| \leq \varepsilon
\end{equation}
with probability at least  $1-\exp(-cm)$ for all $\vz \in\Omega$.
\end{corollary}
\Proof
   We shall give the proof in the Appendix.
\eproof

\begin{corollary}\label{coro:unidenomn}
     Suppose $\chi(t) : \R_+ \to \R $ is a Lipschitz function with Lipschitz constant L and $\supp(\chi) \subset [0,1]$. For any $\delta, \epsilon >0$, if $m\ge Cn$ then with probability at least $1-\exp(-c  m)$ we have
\begin{equation*}  
      \Biggl|\frac{1}{m} \sum_{i=1}^m  \xkh{\frac{\abs{\va_i^\T \vz_1}}{\abs{\va_i^\T \vx}}\chi\Big(\frac{\abs{\va_i^\T \vz_1}}{\abs{\va_i^\T \vx}}\Big) - \frac{\abs{\va_i^\T \vz_2}}{\abs{\va_i^\T \vx}}\chi\Big(\frac{\abs{\va_i^\T \vz_2}}{\abs{\va_i^\T \vx}}\Big) }\Biggr| \leq (L+1)\cdot  \xkh{\norm{\vz_1-\vz_2}/\delta + \delta + \epsilon/\delta + \epsilon}
\end{equation*}
for any fixed $\vz_1,\vz_2 \in \R^d$. Here $c>0$, $C>0$ depend on ($L$, $\delta$, $\epsilon$).
\end{corollary}
\Proof
   We shall give the proof in the Appendix.
\eproof

\subsection{Step 3}  
To prove the main result we will need to estimate several integrals involving the gradient and directional 2nd order derivatives of $F(\vz)$ for $m \geq Cn$ for some $C>0$. We shall show that for a given $\vx\neq 0$, with high probability the Smoothed Amplitude Flow loss function $F(\vz)$ is strictly convex on a small neighborhood of $\pm \vx$. When $\vz$ is not too close to being orthogonal to $\vx$, with high probability $F(\vz)$ has no critical point. Finally, for $\vz$ close to being orthogonal to $\vx$, we show that with high probability any critical point is a saddle point with a negative directional curvature. We shall present these in Section \ref{Proof-main} and the Appendix. 
 
Here ``with high probability'' means there exist constants $c, \delta>0$ such that the probability is at least $1- c\exp (-\delta m) $. We shall make it more explicit in the next section.

\section{Proof of the main results} \label{Proof-main}

\subsection{Notation and General Assumptions}  Now for $\vx\neq 0$ and standard Gaussian random vectors $\{\va_i\}_{i=1}^m$, recall that $F(\vz)$ a continuously differentiable function given by
\begin{equation} \label{mo:mymodelsec}
       F(\vz)=\frac{1}{2m}\sum_{i=1}^m \xkh{\gamma\xkh{\frac{\va_i^\T \vz}
      {\va_i^\T \vx}}-1}^2  \cdot \abs{\va_i^\T \vx}^2,
\end{equation}
with  $\gamma(t)$ being defined in (\ref{eq:gamma}) as
$$
\gamma(t):=\left \{ \begin{array}{cl}
     \abs{t}, &  \abs{t} > \beta;  \\
    \frac{1}{2\beta} t^2+\frac{\beta}{2}, &  \abs{t} \leq \beta.
\end{array} \right. 
$$
We shall denote $f_i(\vz) = \frac{1}{2}\left(\gamma\xkh{\frac{\va_i^\T \vz}
      {\va_i^\T \vx}}-1\right)^2  \cdot \abs{\va_i^\T \vx}^2$, where $\left(\gamma\left(\frac{\va_i^\T \vz}{\va_i^\T \vx}\right)-1\right)^2\cdot \abs{\va_i^\T \vx}^2=  \abs{\va_i^\T \vz}^2$ if $\va_i^\T \vx=0$ . So $F(\vz) = \frac{1}{m}\sum_{i=1}^m f_i(\vz)$. Set
\begin{equation}  \label{eq:Psi}
    \Psi(u,v) = \frac{1}{2} \left(\gamma\Bigl(\frac{u}{v}\Bigr)-1\right)^2 v^2 \quad\mbox{if}~~v \neq 0, 
    \quad\mbox{and}\quad  \Psi(u,0) = \frac{1}{2} u^2. 
\end{equation}
Then $f_i(\vz)= \Psi(\va_i^\T\vz, \va_i^\T\vx)$, and $\nabla f_i(\vz) = \Psi_u(\va_i^\T\vz, \va_i^\T\vx)\va_i$, where 
\begin{equation}  \label{eq:partial_Psi}
    \Psi_u(u,v) = \left(\gamma\Bigl(\frac{u}{v}\Bigr)-1\right)\,\gamma'\Bigl(\frac{u}{v}\Bigr)\,v
        = \left \{ \begin{array}{ll}
             \sgn(u) (|u|-|v|), &  \abs{u} > \beta |v|;  \\
    \frac{1}{2\beta^2}\frac{u^3}{v^2} + (\frac{1}{2}-\frac{1}{\beta})u, &  \abs{u} \leq \beta |v|. \end{array} \right.
\end{equation}
The following summarizes bounds and Lipschitz property of $\Psi_u$:
\begin{equation} \label{eq:Psi_bound}
    |\Psi_u(u,v)| \leq |u|+|v|,~~~\Psi_u(u,v)u \geq u^2- |uv|, 
\end{equation}
and
\begin{equation} \label{eq:Psi_Lip}
    |\Psi_u(u_1,v)-\Psi_u(u_2,v)| \leq \max (1, |2-1/\beta|)\cdot |u_1-u_2|.  
\end{equation}
These properties are easy to check, and we delay the proofs in the appendix.

In the rest of the paper we shall adopt the following notation and specify some assumptions without loss of generality.
\vspace{3mm}

\noindent
\begin{itemize}
\item[(A1)]~~Since  $F(\vz)=F(-\vz)$, to solve the SAF model here we may without loss of generality consider solving the SAF model on the half space $\nj{\vz,\vx} \geq 0$. Denote $\sigma =\sigma(\vz):=\nj{\vz,\vx}/\|\vz\|\|\vx\| \geq 0$ and $\tau =\tau(\vz):= \sqrt{1-\sigma^2}$. Furthermore, we shall write
\begin{equation}  \label{eq:orthogonal}
       \frac{\vz}{\|\vz\|} = \sigma \frac{\vx}{\|\vx\|} + \tau \vw
\end{equation}
where $ \vw \perp \vx$ and $\|\vw\|=1$.

\item[(A2)]~~Denote $\lambda = \beta/\|\vz\|$. Two quantities that appear often in the paper are $\mu_+=\mu_+(\lambda,\sigma)$ and $\mu_-=\mu_-(\lambda,\sigma)$ given by
\begin{align*}
   \mu_+^2  & :=1+\frac{(\sigma+\lambda)^2}{\tau^2} = \frac{1}{\tau^2}(1+\lambda^2+2\sigma\lambda), \\ 
   \mu_-^2  & :=1+\frac{(\sigma-\lambda)^2}{\tau^2} = \frac{1}{\tau^2}(1+\lambda^2-2\sigma\lambda).
\end{align*} 

\item[(A3)]~~We may of course without loss of generality assume that $\|\vx\|=1$. Let $\va \sim N(0,I_n)$ be a standard Gaussian vector in $\Rn$. From the orthogonal decomposition $\vz/\|\vz\| = \sigma \vx + \tau \vw$ in (\ref{eq:orthogonal}), define $U = \va^\T\vz/\|\vz\|, V= \va^\T\vx$ and $W=\va^\T\vw$. Then $U,V,W \sim N(0,1)$ and $V, W$ are independent. Let $A=A(\lambda)$ denote the event
$$
     A = A(\lambda) := \left\{\va:~ |\va^\T\vz| \leq \beta |\va^\T\vx|\right\}
                     = \left\{\va:~ |U| \leq \lambda |V|\right\}.
$$
The notations given here will be used extensively in the next subsection where we compute various expectations. 
\end{itemize}

\begin{definition}
    We say that a property ${\mathcal P}$ holds with {\em high probability for $m$} if there exist some $c, \delta>0$ such that ${\mathcal P}$ holds with probability at least $1-c \exp(-\delta m)$.
\end{definition}

\subsection{Expectations}

Following the notations in (A3) above, we first observe that the condition $|U| \leq \lambda |V|$ is equivalent to the event $\{|\sigma V +\tau W|\leq \lambda|V|\}$, and hence
\begin{equation} \label{eq:AW}
     A= \Bigl\{\tau^{-1}(-\lambda -\sigma\, \sgn(V))|V| \leq W \leq \tau^{-1}(\lambda -\sigma \,\sgn(V))|V|\Bigr\}.
\end{equation}
This is used to prove the following crucial proposition.

\begin{prop}  \label{lem:dlambda}
       Let  $G(\lambda) :=\E[g(U,V)\1_A]$ where $g(t,s)$ is continuous. Then
\begin{align*}  
   \frac{dG}{d\lambda} & =\frac{1}{ 2\pi\tau}\int_0^\infty\left(g(-\lambda v,v)+g(\lambda v,-v)\right)ve^{-\frac{1}{2}\mu_+^2v^2}dv \\
    &\mathrel{\phantom{=}}  +  \frac{1}{ 2\pi\tau}\int_0^\infty\left(g(\lambda v,v)+g(-\lambda v,-v)\right)ve^{-\frac{1}{2}\mu_-^2v^2}dv. 
\end{align*}
\end{prop}
\Proof
          We shall prove this lemma in the appendix.
\eproof

\begin{corollary}  \label{coro:dlambda1}
       Assume that $g(t,s) = |t|^p|s|^q$ in Proposition \ref{lem:dlambda} where $p+q \geq 0$. Then 
\begin{equation*}
   G'(\lambda) = \frac{\lambda^p}{\pi\tau}\left(\mu_-^{-(p+q+2)} +\mu_+^{-(p+q+2)}\right)
       \int_0^\infty t^{p+q+1}e^{-\frac{1}{2}t^2}dt.
\end{equation*}
In particular, if $p+q=2$ then $G'(\lambda) = \frac{2\lambda^p}{\pi\tau}(\mu_-^{-4} +\mu_+^{-4})$.
\end{corollary}
\Proof
   This is a straightforward application of Proposition \ref{lem:dlambda}. Observe that for $p+q=2$ the integral $\int_0^\infty t^3e^{-\frac{1}{2}t^2}dt =2$.
\eproof

\begin{corollary}  \label{coro:dlambda2}
       Assume that $g(t,s) = {\rm sgn} (ts)|t|^p|s|^q$ in Lemma \ref{lem:dlambda} where $p+q \geq 0$. Then 
\begin{equation*}
   G'(\lambda) = \frac{\lambda^p}{\pi\tau}\left(\mu_-^{-(p+q+2)} -\mu_+^{-(p+q+2)}\right)
       \int_0^\infty t^{p+q+1}e^{-\frac{1}{2}t^2}dt.
\end{equation*}
Hence $G(\lambda) \geq 0$. In particular, if $p+q=2$ then $G'(\lambda) = \frac{2\lambda^p}{\pi\tau}(\mu_-^{-4} -\mu_+^{-4})$.
\end{corollary}
\Proof
   Same as the previous corollary.
\eproof

\begin{lemma}  \label{lem:EUV-VW}
\begin{align*}
   \E[|UV|] & =  \frac{2}{\pi}\xkh{\tau+\sigma\arctan{\frac{\sigma}{\tau}}}, \\
   \E[\sgn(UV)V^2)] &= \frac{2}{\pi}\xkh{\tau\sigma +\arctan{\frac{\sigma}{\tau}}}.
\end{align*}
\end{lemma}
\Proof
   We leave the proof to Appendix.
\eproof

\subsection{Non-Vanishing Gradient}
In this subsection we evaluate the gradient of $F(\vz)$. Our ultimate goal is to establish a region on which $\nabla F(\vz)$ does not vanish.

\begin{lemma}  \label{lem:non-vanishing-gradfZ}
       Assume that $0<\beta<1$ and $\|\vx\|=1$. For any $\delta_0>0$ there exist $C,\varepsilon_0>0$ such that with high probability for $m \geq Cn$,
$$
      \nj{\nabla F(\vz),\vz} \geq \varepsilon_0 \|\vz\|^2
$$
for all {  $\|\vz\| \geq \frac{2}{\pi}\xkh{\tau+\sigma\arctan{\frac{\sigma}{\tau}}}+\delta_0$.}
\end{lemma}
\Proof
{
    We have $\nabla F(\vz) = \frac{1}{m} \sum_{i=1}^m \nabla f_i(\vz)$ and  $\nj{\nabla f_i(\vz),\vz} = \Psi_u (\va_i^\T \vz, \va_i^\T \vx)(\va_i^\T\vz)$. By \eqref{eq:Psi_bound} we have
 $$
       \Psi_u (\va_i^\T \vz, \va_i^\T \vx)(\va_i^\T\vz) \geq (\va_i^\T\vz)^2 - |(\va_i^\T\vz)(\va_i^\T\vx)|.
$$
Set $U_i = \va_i^\T\vz/\|\vz\|, V_i = \va_i^\T\vx$. Then from Lemma \ref{lem:EUV-VW} we have
$$
    \E[|U_iV_i|] = \frac{2}{\pi}\xkh{\tau+\sigma\arctan{\frac{\sigma}{\tau}}}.
$$
It follows that 
$$
    \frac{1~}{\|\vz\|^2}\E[\nj{\nabla f_i(\vz),\vz}] \geq 1 - \frac{1}{\|\vz\|}\E[|U_iV_i|] \geq 1- \frac{1}{\|\vz\|}\cdot \frac{2}{\pi}\xkh{\tau+\sigma\arctan{\frac{\sigma}{\tau}}}. 
$$
Note that $\frac{1~}{\|\vz\|^2}\nj{\nabla f_i(\vz),\vz}$ is continuous satisfying the condition in Corollary \ref{coro:union}. It follows that for any $\delta_0>0$ there exists $C, \ep_0>0$ such that with high probability for $m \geq Cn$ we have
$$
   \frac{1~}{\|\vz\|^2}\nj{\nabla F(\vz),\vz} =\frac{1}{m}\sum_{i=1}^m \frac{1~}{\|\vz\|^2}\nj{\nabla f_i(\vz),\vz} \geq \ep_0
$$
for all $\|\vz\| \geq \frac{2}{\pi}\xkh{\tau+\sigma\arctan{\frac{\sigma}{\tau}}}+\delta_0$ (This region is a compact set with respect to $\lambda=\beta/\norms{\vz}$). The lemma is proved.
\eproof

\begin{lemma}  \label{lem:negativeGradfX}
       Assume that $0<\beta\leq \frac{1}{2}$ and $\|\vx\|=1$. For any $\ep_0, \sigma_0>0$ there exist $C,\epsilon>0$ such that with high probability for $m\geq Cn$,  we have $\nj{\nabla F(\vz),\vx} <-\epsilon$ for all $\vz$ such that  $\ep_0 \leq\|\vz\| \leq 1$ and $\sigma_0 \leq \sigma \leq 1-\sigma_0$.
\end{lemma}
\Proof  We have $\nj{\nabla F(\vz),\vx} = \frac{1}{m} \sum_{i=1}^m  \Psi_u (\va_i^\T \vz, \va_i^\T \vx)(\va_i^\T\vx)$. Let $\va \sim N (0, I_n)$ be standard Gaussian. Then $\E[\nj{\nabla F(\vz),\vx}] = \E[ \Psi_u (\va^\T \vz, \va^\T \vx)(\va^\T\vx)]$. Set 
$$
       g(\vz) = \frac{1}{\|\vz\|}\E[\nj{\nabla F(\vz),\vx}] =\frac{1}{\|\vz\|}\E[ \Psi_u (\va^\T \vz, \va^\T \vx)(\va^\T\vx)]. 
$$
Using the notation $U, V, W$ and $\lambda = \beta/\|\vz\|$ defined in (A3) and expanding out $\Psi_u$ via \eqref{eq:partial_Psi}, it yields 
$$
   g(\vz) = \sigma - \frac{\lambda}{\beta} \E[\sgn(UV)V^2\1_{A^c}] +\frac{1}{2\lambda^2}\,\E[U^3V^{-1})\1_{A}]
             - \Bigl(\frac{1}{2}+\frac{1}{\beta}\Bigr)\E[UV\1_A]
$$
where $A:=\{|U| \leq \lambda |V|\}$. Note that $g(\vz)$ now depends on $\lambda$ and $\sigma$ as $\vz$ is completely determined by $\lambda$ and $\sigma$ once $\beta$ is fixed. Set $B(\lambda, \sigma) := \frac{2}{\pi\tau}(\mu_-^{-4}-\mu_+^{-4})$. By Corollary \ref{coro:dlambda2} we have
\begin{align*}
    \frac{\partial g}{\partial\lambda}
        &= - \frac{1}{\beta} \E[\sgn(UV)V^2\1_{A^c}] + \frac{\lambda}{\beta}\cdot B(\lambda,\sigma)
              - \frac{1}{\lambda^3}\,\E[U^3V^{-1})\1_{A}]  \\
         & \mathrel{\phantom{=}} +\frac{1}{2\lambda^2}\cdot \lambda^3 B(\lambda,\sigma)  - \Bigl(\frac{1}{2}+\frac{1}{\beta}\Bigr)  \lambda B(\lambda,\sigma) \\
        & = - \frac{1}{\beta} \E[\sgn(UV)V^2\1_{A^c}] - \frac{1}{\lambda^3}\,\E[U^3V^{-1})\1_{A}].
\end{align*}
Also by Corollary \ref{coro:dlambda2},  we know
\[
\E[U^3V^{-1})\1_{A}]=\frac{2}{\pi \tau} \int_0^\lambda t^3 \xkh{\mu_-^{-4}+\mu_+^{-4}} ~dt \geq 0.
\]
 Furthermore, note that $A^c=\{|U| \geq \lambda |V|\} = \{|V| < \lambda^{-1} |U|\}$. By switching the role of $U$ and $V$ we have
$$
      \E[\sgn(UV)V^2\1_{A^c}] = \E[\sgn(UV)U^2\1_{A'}] =\frac{2}{\pi \tau} \int_0^{\lambda^{-1}} t^2 \xkh{\mu_-^{-4}-\mu_+^{-4}} ~dt\geq 0
$$
where $A' := \{|U| \leq \lambda^{-1} |V|\}$. { Therefore $g$ is a decreasing function in $\lambda$, which means that $g$ is increasing with respect to $\norms{\vz}$}.

We would like to prove that $g(\vz) <0$ for $\ep_0 \leq \|\vz\| \leq 1$ and $\sigma_0 \leq \sigma \leq 1-\sigma_0$. To do so we only need to show $g(\vz) <0$ for $\|\vz\|=1$, i.e. $\lambda=\beta$ and $\sigma_0 \leq \sigma \leq 1-\sigma_0$. Now, note that 
\begin{align*}
      \E[\sgn(UV)V^2\1_{A^c}] &=  \E[\sgn(UV)V^2] -\E[\sgn(UV)V^2\1_{A}]  \\
         &= \frac{2}{\pi}\Bigl(\tau\sigma +\arctan{\frac{\sigma}{\tau}}\Bigr)  -\E[\sgn(UV)V^2\1_{A}].  
\end{align*}
Thus for $\lambda=\beta$, again applying Corollary \ref{coro:dlambda2} we obtain
\begin{equation} \label{eq:gradfdX-int}
    g(\vz) = \sigma -\frac{2}{\pi}\Bigl(\tau\sigma +\arctan{\frac{\sigma}{\tau}}\Bigr) + 
        \int_0^\beta \left(1+  \frac{t^3}{2\beta^2}- \Big(\frac{1}{2}+\frac{1}{\beta}\Bigr)t
       \right)B(t,\sigma)\,dt .                   
\end{equation}
To establish $g(\vz)<0$ here, we observe that 
$$
   \frac{\partial g(\vz)}{\partial \beta} = \int_0^\beta \left(\frac{1}{\beta}\Bigl(\frac{t}{\beta}\Bigr)\,-\,
         \Bigl(\frac{t}{\beta}\Bigr)^3 \right)B(t,\sigma)\,dt  >0.
$$
Thus $g(\vz)$ is increasing with respect to $\beta$. Hence, it suffices to show that $g(\vz)<0$ for $\beta = 1/2$. Expanding $B(t, \sigma)$ yields
$$
    B(t, \sigma) = \frac{16\tau^3\sigma t (1+t^2)}{\pi \left((1 + t^2)^2 -4\sigma^2 t^2\right)^2} 
                   =: \tau^3\sigma \,Q(t, \sigma),
$$
where $Q(t,\sigma) := \frac{16 t (1+t^2)}{\pi \left((1 + t^2)^2 -4\sigma^2 t^2\right)^2}$. It follows that for $\beta=1/2$ we have
\begin{equation} \label{eq:gradfdX-int2}
   \frac{1}{\sigma \tau^3} g(\vz) = \frac{1}{\tau^3}\left(1 -\frac{2}{\pi}\Bigl(\tau  +\frac{1}{\sigma}\arctan{\frac{\sigma}{\tau}}\Bigr)\right) + 
        \int_0^{1/2} \left( 1+  2t^3-2.5 t
       \right)Q(t,\sigma)\,dt .                   
\end{equation}
Clearly $Q(t,\sigma) \leq Q(t, 1)$. Next, integrating rational functions by partial fractions, we can obtain
\begin{eqnarray*}
      \int_0^{1/2} \left( 1+  2t^3-2.5 t
       \right)Q(t,1)~dt &=  & \frac{8}{\pi}\int_0^{1/2} \frac{(4t^3-5t+2)(t^2+1)t}{(t^2-1)^4}~dt \\
      &=& \frac{4}{\pi}\xkh{\frac{35}{27}-\ln 3} < 0.26.
\end{eqnarray*}
Meanwhile, the term before the integral in \eqref{eq:gradfdX-int2} is a function of $\sigma$ and  it is decreasing. Indeed,  by Corollary \ref{coro:dlambda2} where we set $\lambda=\infty$, we have
\begin{eqnarray*}
P(\sigma) &:= & \frac{1}{\tau^3}\left(1 -\frac{2}{\pi}\Bigl(\tau  +\frac{1}{\sigma}\arctan{\frac{\sigma}{\tau}}\Bigr)\right) \\
&=&  \frac{1}{\tau^3\sigma} \E\zkh{UV-\sgn(UV)V^2}\\
&=&   \frac{1}{\tau^3\sigma}  \int_0^\infty (t-1)B(t,\sigma) ~dt\\
&=&  \frac{16}{\pi}  \int_0^\infty  \frac{(t-1)(1+t^2)t}{\zkh{(1+t^2)^2-4t^2\sigma^2}^2}~dt.
\end{eqnarray*}
Making a substitution $t=\frac{1}{u}$, we obtain
\[
 \int_1^\infty  \frac{(t-1)(1+t^2)t}{\zkh{(1+t^2)^2-4t^2\sigma^2}^2}~dt= -\int_0^1  \frac{(u-1)(1+u^2)u^2}{\zkh{(1+u^2)^2-4u^2\sigma^2}^2}~du.
\]
It gives 
\[
P(\sigma)=-\frac{16}{\pi}  \int_0^1  \frac{(1-t)^2(1+t^2)t}{\zkh{(1+t^2)^2-4t^2\sigma^2}^2}~dt,
\]
which is decreasing with respect to $\sigma$. Thus the maximum of $P(\sigma)$ is achieved at $\sigma=0$ and we have\footnote{In the appendix, we shall
prove a stronger inequality for  the monotonicity of the function 
$f_0(\tau)=
 \frac 1 {\tau^2} ( 1- \frac 2{\pi}  (\tau+\frac 1 {\sigma} \arctan \frac{\sigma}{\tau} ))$.
}
\begin{eqnarray*}
 \frac{1}{\tau^3}\left(1 -\frac{2}{\pi}\Bigl(\tau  +\frac{1}{\sigma}\arctan{\frac{\sigma}{\tau}}\Bigr)\right)& \le & \lim_{\sigma \to 0} \left( 1 -\frac{2}{\pi} \Bigl(1  +\frac{\arctan\sigma}{\sigma} \Bigr)\right) \\
 &=& 1-\frac{4}{\pi}\leq -0.27.
\end{eqnarray*}
Thus we have shown by combining the two estimates that $\frac{1}{\sigma \tau^3} g(\vz) <-0.01$. So $g(\vz) < -\delta_0$ for some $\delta_0>0$.

To show that $\nj{\nabla F(\vz),\vx} <-\epsilon$ with high probability for $m \geq Cn$ for all $\ep_0 \leq \|\vz\| \leq 1$ and $\sigma_0 \leq \sigma \leq 1-\sigma_0$, set $U_i = \va_i^\T\vz$, $V_i = \va_i^\T\vx$. Then $\frac{1}{\|\vz\|}\nj{\nabla f_i(\vz),\vx}$ is a continuous function in $U_i$ and $\lambda = \beta/\|\vz\|$, where $\beta \leq\lambda \leq \beta/\sigma_0$. One can easily check that the conditions of Corollary \ref{coro:union} are satisfied by applying \eqref{eq:Psi_Lip} . The lemma now follows.
\eproof

\begin{lemma}  \label{lem:negativeGradfX-sigma1}
       Assume that $0<\beta<1$ and $\|\vx\|=1$. For any $\ep_0>0$ there exist $C,  \sigma_0,\epsilon>0$ such that with high probability for $m\geq Cn$,  we have $\nj{\nabla F(\vz),\vx} <-\epsilon$ for all $\vz$ such that  $\ep_0 \leq\|\vz\| \leq 1-\ep_0$ and $\sigma \geq 1-\sigma_0$.
\end{lemma}
\Proof  Again we start from $\nj{\nabla F(\vz),\vx} = \frac{1}{m} \sum_{i=1}^m  \Psi_u (\va_i^\T \vz, \va_i^\T \vx)(\va_i^\T\vx)$. Let $\va \sim N (0, I_n)$ be standard Gaussian. Then $\E[\nj{\nabla F(\vz),\vx}] = \E[ \Psi_u (\va^\T \vz, \va^\T \vx)(\va^\T\vx)]$. First we consider $\sigma=1$, for which $\vz = \|\vz\|\vx$ and $\va^\T \vz = \|\vz\|\va^\T \vx$.  Set $\lambda = \beta/\|\vz\|$ as usual. One can easily check via \eqref{eq:partial_Psi} that 
$$
   \Psi_u (\va^\T \vz, \va^\T \vx)(\va^\T\vx) = \left\{\begin{array}{ll} 
                           (\|\vz\|-1) (\va^\T\vx)^2, & \|\vz\|>\beta; \\ 
                           \frac{\|\vz\|^3}{2\beta^2}(\va^\T\vx)^2 + (\frac{1}{2}-\frac{1}{\beta})\|\vz\|(\va^\T\vx)^2, & \|\vz\|\leq\beta. 
  \end{array}   \right.
$$
Thus $\Psi_u (\va^\T \vz, \va^\T \vx)(\va^\T\vx) \leq -\delta_0 (\va^\T\vx)^2$ for some $\delta_0>0$. It follows that 
$\E[\nj{\nabla F(\vz),\vx}] = \E[ \Psi_u (\va^\T \vz, \va^\T \vx)(\va^\T\vx)] \leq -\delta_0$. Now by continuity $\E[\nj{\nabla F(\vz),\vx}]  \leq -\delta_1$ for some $\sigma_0,\delta_1>0$ for all $\ep_0 \leq\|\vz\| \leq 1-\ep_0$ and $\sigma \geq 1-\sigma_0$.

To show that $\nj{\nabla F(\vz),\vx} <-\epsilon$ with high probability for $m \geq Cn$ for all $\ep_0 \leq \|\vz\| \leq 1-\ep_0$ and $\sigma \geq 1-\sigma_0$, Observe that $\Psi_u (\va^\T \vz, \va^\T \vx)(\va^\T\vx)$ is a Lipschitz continuous function of $\vz$ on this region. The conditions of Corollary \ref{coro:union} are met by applying \eqref{eq:Psi_Lip} . The lemma now follows.
\eproof

\subsection{Negative Directional Curvature and Strong Convexity}

To cover the remaining region we evaluate the { second order directional} derivatives of the target function $F(\vz)$. First, the Hessian of $F$ is given by  $\nabla^2 F(\vz) = \frac{1}{m} \sum_{i=1}^m \nabla^2_\vz \Psi(\va_i^\T\vz, \va_i^\T\vx)$ for those points $\vz$ at which $F(\vz)$ has second derivative. For any $\vv\in\Rn$ we have $D_\vv^2 F(\vz) = \frac{1}{m} \sum_{i=1}^m D_\vv^2 \Psi(\va_i^\T\vz, \va_i^\T\vx)$ where $D_\vv^2$ is with respect to the variable $\vz$. It is easy to check that 
\begin{equation} \label{eq:hessian}
     \nabla^2 \Psi(\va_i^\T\vz, \va_i^\T\vx) = \va_i \va_i^\T+ \frac{3}{2\beta^2} \frac{\abs{\va_i^\T \vz}^2}{\abs{\va_i^\T \vx}^2}\cdot \1_{R_i}\va_i \va_i^\T    -
(\frac{1}{2}+\frac{1}{\beta})\cdot \1_{R_i} \va_i \va_i^\T
\end{equation}
{ for all points at  which the Hessian is well-defined, i.e. $\vz \not\in \partial R_i$ with $R_i:=\{|\va_i^\T\vz| < \beta |\va_i^\T\vx|\}$. Here, we use the notation $\partial R$ to denote the boundary points of the set $R$}.
Let 
\begin{equation} \label{eq:phi}
      \phi(t) := 1+\frac{3}{2\beta^2}t^2\1_{\{|t| < \beta\}} - (\frac{1}{2}+\frac{1}{\beta})\cdot \1_{\{|t| < \beta\}}.
\end{equation}
Then $ \nabla^2 \Psi(\va_i^\T\vz, \va_i^\T\vx)  = \phi\bigl(\frac{\va_i^\T\vz}{\va_i^\T\vx}\bigr)\va_i^\T\va_i$  for  $\vz \not\in \partial R_i$. On the other hand, one can check that the 2nd order directional derivative of $\Psi(\va_i^\T \vz, \va_i^\T \vx)$ with respect to $\vz$ is well-defined for $\vz\in\partial R_i$, which is $D_{\vv}^2 \Psi(\va_i^\T \vz, \va_i^\T \vx)  = { (\va_i^\T \vv)^2} $ if $(\va_i^\T \vz)(\va_i^\T \vv)> 0$ and $D_{\vv}^2 \Psi(\va_i^\T \vz, \va_i^\T \vx)  = {(2-1/\beta)(\va_i^\T \vv)^2 }$ if $(\va_i^\T \vz)(\va_i^\T \vv)\leq 0$.  In summary, we have
\begin{align}  
    D_{\vv}^2 \Psi(\va_i^\T \vz, \va_i^\T \vx) &=(\va_i^\T \vv)^2 
              + \frac{3}{2\beta^2} \frac{\abs{\va_i^\T \vz}^2} {\abs{\va_i^\T \vx}^2}\cdot  \abs{\va_i^\T \vv}^2   \1_{R_i} 
    -(\frac{1}{2}+\frac{1}{\beta}) (\va_i^\T \vv)^2    \1_{R_i} 
    + \Gamma_i(\vz, \vv) \nonumber \\
     & =  \phi\Bigl(\frac{\va_i^\T\vz}{\va_i^\T\vx}\Bigr)\,(\va_i^\T \vv)^2 
     + \Gamma_i(\vz, \vv) \label{eq:D2v}
\end{align}
for {\em all} $\vz$, with
$$
    \Gamma_i(\vz,\vv) :=  (q_i-1)\, (\va_i^\T \vv)^2
             \1_{\{ \abs{\va_i^\T \vz} = \beta \abs{\va_i^\T \vx} \}} \leq 0
$$
where $q_i=1$ if $(\va_i^\T \vz)(\va_i^\T \vv)> 0$ and $q_i = 2-1/\beta $ if $(\va_i^\T \vz)(\va_i^\T \vv)\leq 0$. 

\begin{lemma}  \label{lem:negativeD2x}
       Assume that $0<\beta \leq 3/4$ and $\|\vx\|=1$. There exist  $C,\sigma_0, \delta_0,\ep_0>0$ such that with high probability for $m\geq Cn$ we have $D_\vx^2F(\vz) < -\ep_0$ for all $\vz$ such that $\|\vz\| \leq 1+\delta_0$ and $\sigma \leq \sigma_0$, as well as all $\vz\in B_{\delta_0}(0)$. 
\end{lemma}
\Proof  First we assume $\sigma=0$, and prove $\E[D_\vx^2F(\vz)] < -5\ep_0$ if $\|\vz\| \leq 1+\delta_0$ for some $\ep_0, \delta_0>0$.  Let $\va \sim N (0, I_n)$ be standard Gaussian.  Set
$$
     G(\vz) := (\va^\T\vx)^2 + \frac{3}{2\beta^2} (\va^\T\vz)^2   \1_{R} 
    -\Bigl(\frac{1}{2}+\frac{1}{\beta}\Bigr) (\va^\T\vx)^2 \1_{R} = 
    \phi\Bigl(\frac{\va^\T\vz}{\va^\T\vx}\Bigr) (\va^\T\vx)^2
$$
where $R:=\{|\va^\T\vz| < \beta |\va^\T\vx|\}$ and $\phi(t)$ is given in \eqref{eq:phi}. From \eqref{eq:D2v} with $\vv=\vx$  we have
$D_{\vx}^2 \Psi(\va^\T \vz, \va^\T \vx)  \leq G(\vz)$. 
As before let $\lambda = \beta/\|\vz\|$, $U=\va^\T\vz/\|\vz\|$ and $V= \va^\T\vx$. We have
$$
    \E[G(\vz)] = 1 +  \frac{3}{2\lambda^2}\E[U^2\1_A] 
                - \Bigl(\frac{1}{2}+\frac{1}{\beta}\Bigr)\E[V^2\1_A] =: g(\lambda)
$$
where $A =\{|U| \leq \lambda |V|\}$. Observe that by Corollary \ref{coro:dlambda1},
$$ 
    g'(\lambda) = -\frac{3}{\lambda^3}\E[U^2\1_A] +\frac{3}{\pi\tau}(\mu_+^{-4} +\mu_-^{-4}) -\Bigl(\frac{1}{2}+\frac{1}{\beta}\Bigr)\frac{2}{\pi\tau}(\mu_+^{-4} +\mu_-^{-4}) <0.
$$
So $g(\lambda)$ is decreasing with respect to $\lambda$, which also means $\E[G(\vz)]$ is increasing with respect to $\|\vz\|$. We show that $\E[G(\vz)]<-5\ep_0$ for $\|\vz\|\leq 1$, and it suffices to show this for $\|\vz\|=1$, i.e.  $g(\lambda) < -5\ep_0$. 

Since $\sigma=0$ (and hence $\tau=1$) we have simple closed form solution for $g(\lambda)$, which by Corollary \ref{coro:dlambda1} is
\begin{align*}
     g(\lambda) &= 1 +  \frac{3}{2\lambda^2}\cdot \frac{4}{\pi}\int_0^\lambda \frac{t^2}{(1+t^2)^2}\,dt
                - \Bigl(\frac{1}{2}+\frac{1}{\beta}\Bigr)\cdot \frac{4}{\pi}\int_0^\lambda \frac{1}{(1+t^2)^2}\,dt  \\
        &=1 +  \frac{3}{\pi\lambda^2}\Bigl(\arctan \lambda -\frac{\lambda}{1+\lambda^2}\Bigr)
                - \frac{\beta+2}{\pi\beta}\Bigl(\arctan\lambda +\frac{\lambda}{1+\lambda^2}\Bigr).
\end{align*}
Since  $\|\vz\|=1$, it follows that
$$
    g(\lambda) = 1 -\frac{3+\beta^2+2\beta}{\pi(1+\beta^2)\beta}+  \frac{3-\beta^2-2\beta}{\pi\beta^2}\cdot\arctan\beta \quad : = g_0(\beta).
$$
The graph of this function $g_0(\beta)$ is shown in Figure \ref{figure:g_0}; we check that  $g_0(\beta) < -0.03$ for all $\beta \in (0,\frac{3}{4}]$.
Taking $\ep_0=0.006$ we obtain the desired result. Since $g(\lambda)$ is continuous, for a sufficiently small $\delta_0>0$ we have $g(\lambda) <-4\ep_0$ for all $\|\vz\| \leq 1+\delta_0$. 

\begin{figure}[H]
\centering
     \includegraphics[width=0.5\textwidth]{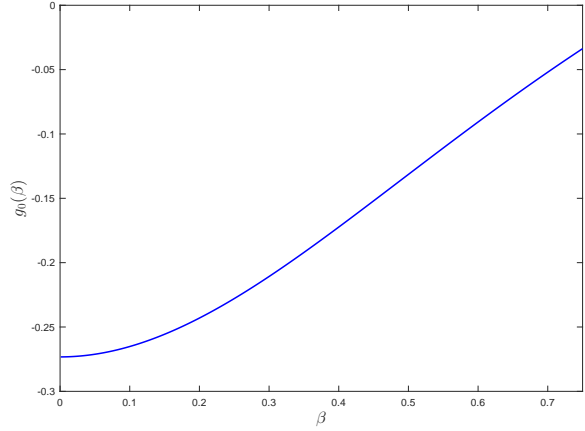}
\caption{ $g_0(\beta)$.}
\label{figure:g_0}
\end{figure}

To prove the lemma for the region $\|\vz\| \leq 1+\delta_0$ and $\sigma \leq \sigma_0$, we need to use Lipschitz condition and Corollary \ref{coro:union}. Our challenge is that the function $\phi(t)$ is discontinuous with a jump discontinuity at $|t|=\beta$. To get around this problem we smooth out the function by introducing 
$$
    H_p(u,v) := v^2 + \frac{3}{2\beta^2} u^2   \1_{S} 
    -\Bigl(\frac{1}{2}+\frac{1}{\beta}\Bigr) v^2 \1_{S} 
    +\Bigl(\frac{1}{\beta}-1\Bigr) {\frac{|u|^p}{\beta^p |v|^{p-2}}\1_S}
$$
where $p>0$ and $S=\{(u,v):~|u| < \beta|v|\}$. A key observation is $H_p(u,v)$ is continuous and satisfies the Lipschitz condition { with respect to $u$} in Corollary \ref{coro:union} (here we need the property $|u| < \beta |v|$ on $S$). Clearly 
 $H_p(\va^\T\vz, \va^\T\vx) \geq G(\vz)$.  {Note that 
 \[
 \frac{|u|^p}{\beta^p |v|^{p-2}}\1_S=\xkh{\frac{\abs{u}}{\beta \abs{v}}}^p \abs{v}^2\1_S 
 \;\ra\; 0 \quad  \mbox{as} \quad p\ra \infty.
 \]
 It means that $\E[H_p(\va^\T\vz, \va^\T\vx)] \lra \E[G(\vz)]$ as $p \ra \infty$}. Hence  for $0\leq\|\vz\| \leq 1+\delta_0$, by taking $p=p_0$ sufficiently large we obtain $\E[H_{p_0}(\va^\T\vz, \va^\T\vx)]<-3\ep_0$. Because $\E[H_{p_0}(\va^\T\vz, \va^\T\vx)]$ is a continuous function of $\sigma$, there exists a $\sigma_0>0$ such that $\E[H_{p_0}
(\va^\T\vz, \va^\T\vx)]<-2\ep_0$ for $0<\|\vz\| \leq 1+\delta_0$ and $\sigma \leq \sigma_0$. Finally,
$$
   D_\vx^2 F(\vz) \leq \frac{1}{m} \sum_{i=1}^m H_{p_0}(\va_i^\T\vz, \va_i^\T\vx).
$$
It follows from Lemma \ref{lem:concentration} and Corollary \ref{coro:union} that with high probability for $m \geq Cn$, $D_\vx^2 F(\vz) \leq -\ep_0$ for all $\vz$ with $0 \leq \|\vz\| \leq 1+\delta_0$ and $\sigma \leq \sigma_0$.

To prove the lemma for $\vz \in B_{\delta_0}(0)$ we use virtually identical technique. It is easily seen $\E[H_p(0)] = \E[G(0)] =\frac{1}{2}-\frac{1}{\beta}<0$. By continuity $\E[H_p(\vz)]< -\ep_1$ for $\vz \in B_{\delta_0}(0)$ for some $\delta_0, \ep_1>0$. Now using Lemma \ref{lem:concentration} and Corollary \ref{coro:union} the same argument as in the previous region apply to prove this case of the lemma.   
\eproof

\begin{lemma}  \label{lem:convex}
       Assume that $0<\beta<1$ and $\|\vx\|=1$. There exist  $C, \delta_0,\ep_0>0$ such that with high probability for $m\geq Cn$ we have $D_\vv^2F(\vz) \geq 0.5$ for all $\vz \in B_{\delta_0}(\vx)$ and unit vectors $\vv \in\Rn$. In other words, $F(\vz)$ is strongly convex in a neighborhood of the solution $\vx$. 
\end{lemma}
\Proof Let $\va \sim N (0, I_n)$ be standard Gaussian.  We follow the same strategy of evaluating an expectation and prove the lemma using Lipschitz condition and  Corollary \ref{coro:union}. Again, due to the discontinuity of $\phi(t)$ in \eqref{eq:D2v} we will need to smooth it out. Define an auxiliary function $\chi_1(x)$ for $x\ge 0$ as
\begin{equation*}
\chi_1(x):=\left\{ \begin{array}{ll}
                      1 & \mbox{if} \quad 0\le x \le  \beta, \vspace{1em} \\
                      1+\frac{\beta}{\delta_0}-\frac{1}{\delta_0}x & \mbox{if} \quad \beta \le x \le  \beta+\delta_0, \vspace{1em}\\
                      0 &  \mbox{if} \quad x >\beta+\delta_0,
                      \end{array}
                      \right. 
\end{equation*}
where $\delta_0=\sqrt{\frac{2\beta+\beta^2}{3}}-\beta$. Furthermore, let $\chi_2(x) \in C_c^\infty(\R)$ be a function such that $0\le \chi_2(x) \le 1$ for all $x$, $\chi_2(x) =1$ for $\abs{x} \le 1$ and $\chi_2(x) =0$ for $\abs{x} \ge 2$.
Define $G(u,v,t)$ such that
\begin{align*}
     &G(\va^\T\vz, \va^\T\vx, \va^\T\vv)  \notag \\
     := &\;(\va^\T \vv)^2 
              + \zkh{\frac{3}{2\beta^2} \frac{\abs{\va^\T \vz}^2} {\abs{\va^\T \vx}^2}\cdot  
              \abs{\va^\T \vv}^2  -(\frac{1}{2}+\frac{1}{\beta}) (\va^\T \vv)^2} \chi_1\xkh{ \frac{\abs{\va^\T \vz}} {\abs{\va^\T \vx}}}\chi_2\xkh{ \frac{\abs{\va^\T \vv}} {M_0\abs{\va^\T \vx}}},
\end{align*}
where $M_0$ is a positive constant which will be made sufficiently large.

Note that $\chi_1(x)$ is enlarged from the set $R=\{|\va^\T\vz| < \beta |\va^\T\vx|\}$. Set $\vz=\vx$, Then
$$
    G(\va^\T\vx, \va^\T\vx, \va^\T\vv) 
    = (\va^\T \vv)^2 
              + (\frac{3}{2\beta^2}-\frac{1}{2}-\frac{1}{\beta}) (\va^\T \vv)^2 \cdot \chi_2\xkh{ \frac{\abs{\va^\T \vv}} {M_0\abs{\va^\T \vx}}} \geq (\va^\T \vv)^2.
$$
Hence $\E[G(\va^\T\vx, \va^\T\vx, \va^\T\vv)] \geq 1$. Since $G(\va^\T\vz, \va^\T\vx, \va^\T\vv)$ is continuous with respect to $\vz$ and $\vv$, and $\vv$ is on the unit sphere which is compact, we know that $$\E[G(\va^\T\vz, \va^\T\vx, \va^\T\vv)] \geq 0.7$$
 for $\vz$ in a neighborhood $\vz\in B_{\delta_0}(\vx)$ for all unit vectors $\vv$.
Next, we show $G(\va^\T\vz, \va^\T\vx, \va^\T\vv)$ satisfies the Lipschitz type condition with respect to $\vz, \vv$. To this end, we only need to consider the second term of $G(\va^\T\vz, \va^\T\vx, \va^\T\vv)$.  Note that
\begin{eqnarray*}
L(\va^\T\vz, \va^\T\vx, \va^\T\vv)&:=& \frac{\abs{\va^\T \vz}^2} {\abs{\va^\T \vx}^2}\cdot  \abs{\va^\T \vv}^2  \chi_1\xkh{ \frac{\abs{\va^\T \vz}} {\abs{\va^\T \vx}}}\chi_2\xkh{ \frac{\abs{\va^\T \vv}} {M_0\abs{\va^\T \vx}}} \\
&=& \frac{\abs{\va^\T \vz}^2} {\abs{\va^\T \vx}^2} \chi_1\xkh{ \frac{\abs{\va^\T \vz}} {\abs{\va^\T \vx}}} \cdot  \frac{\abs{\va^\T \vv}^2} {\abs{\va^\T \vx}^2}\chi_2\xkh{ \frac{\abs{\va^\T \vv}} {M_0\abs{\va^\T \vx}}} \cdot \abs{\va^\T \vx}^2 \\
&=:\;& \psi_1\xkh{\frac{\abs{\va^\T \vz}} {\abs{\va^\T \vx}}}  \psi_2\xkh{\frac{\abs{\va^\T \vv}} {\abs{\va^\T \vx}}} \abs{\va^\T \vx}^2.
\end{eqnarray*}
It is obvious that $\psi_1$ and $\psi_2$ are Lipschitz and bound functions. Observe that for any $\vz_1$ and $\vz_2$, we have
\[
\Abs{\psi_1\xkh{\frac{\abs{\va^\T \vz_1}} {\abs{\va^\T \vx}}}-\psi_1\xkh{\frac{\abs{\va^\T \vz_2}} {\abs{\va^\T \vx}}}} \lesssim \frac{\abs{ \va^\T (\vz_1-\vz_2) }}{\abs{\va^\T \vx}}.
\]
Thus, 
\[
\Abs{L(\va^\T\vz_1, \va^\T\vx, \va^\T\vv)-L(\va^\T\vz_2, \va^\T\vx, \va^\T\vv)} \lesssim \abs{ \va^\T (\vz_1-\vz_2) }\abs{\va^\T \vx}.
\]
By Corollary \ref{coro:union} and Corollary \ref{coro:unidenomn}, we know the function $G(\va^\T\vz, \va^\T\vx, \va^\T\vv)$ satisfies the Lipschitz type condition with respect to $\vz, \vv$.

It follows that with high probability for $m \geq Cn$ we have
$$
      \frac{1}{m}\sum_{i=1}^m G(\va_i^\T\vz, \va_i^\T\vx, \va_i^\T\vv) \geq 0.5
$$
for all $\vz\in B_{\delta_0}(\vx)$ and unit vectors $\vv$.   Choosing the constant $M_0$ sufficiently large and  combining with \eqref{eq:D2v} we see that $G(\va_i^\T\vz, \va_i^\T\vx, \va_i^\T\vv) \leq D_\vv^2 \Psi(\va_i^\T\vz, \va_i^\T\vx)$ . Hence with high probability for $m \geq Cn$ we have
$$
    D_\vv^2F(\vz) \geq  \frac{1}{m}\sum_{i=1}^m G(\va_i^\T\vz, \va_i^\T\vx, \va_i^\T\vv) \geq 0.5
$$
for all $\vz\in B_{\delta_0}(\vx)$ and unit vectors $\vv$.
\eproof

\subsection{Putting Things Together}
~~
\noindent

\vspace{2mm}
\noindent
{\bf Proof of Theorem \ref{theo-1.1}.}~~Without loss of generality we shall only examine the region $\sigma \geq 0$, i.e. $\nj{\vz, \vx} \geq 0$. The lemmas we have proved in this section have covered all regions to ensure that with high probability for $m \geq Cn$: (i) In a small neighborhood of  $\vx$ the target function $F(\vz)$ is strongly convex with $\vz=\vx$ being a minimum. (ii) Everywhere else either the gradient of $F(\vz)$ doesn't vanish or $F(\vz)$ has a negative directional curvature. Thus other than $\vx$ the target function $F(\vz)$ has no other local minimum. This proves the theorem.
\eproof

\section{Numerical Experiments}

 The SAF model proposed in this study shows theoretically that any gradient descent algorithm will not get trapped in a local minimum. Here we present numerical experiments to show that the model performs very well with random initial guess.

   We use the following vanilla gradient descent algorithm 
\[
     \vz_{k+1}=\vz_{k}-\mu \nabla F(\vz_{k})
\]
with a random initial guess to minimize the loss function $F(\vz)$ of SAF. The algorithm procedure is as follows:

\begin{algorithm}[H]
\caption{Gradient Descent Algorithm Based on Smoothed Amplitude Flow (SAF)}
\begin{algorithmic}[H]
\Require
Measurement vectors: $a_i\in \R^n, i=1,\ldots,m $; Observations: $y \in \R^m$;  Parameters $\beta$; Step size $\mu$; Tolerance $\epsilon>0$  \\
\begin{enumerate}
\item[1:] Random initial guess $\vz_0\in \Rn$.
\item[2:] For $k=0,1,2,\ldots,$ if $\norm{\nabla F(\vz_{k})} \ge \epsilon $ do
\[
\vz_{k+1}=\vz_{k}-\mu \nabla F(\vz_{k})
\]
\item[3:] End do
\end{enumerate}

\Ensure
The vector $ \vz_T $.
\end{algorithmic}
\end{algorithm}

The performance of our SAF algorithm is conducted via a series of numerical experiments in comparison against WF \cite{WF}, TWF \cite{TWF} and TAF \cite{TAF}. Here, it is worth emphasizing that random initialization is used for our SAF algorithm while all other algorithms have adopted a spectral initialization.  Our theoretical results are for real Gaussian case, but the algorithms can be easily adapted to the complex Gaussian case. In our numerical experiments, the target vector $\vx\in \Rn$ is chosen randomly from the standard Gaussian distribution and the measurement vectors $\va_i, \,i=1,\ldots,m$ are also generated randomly from standard Gaussian distribution.   For the real Gaussian case, the signal $\vx \sim  \mathcal{N}(0,I_n)$ and measurement vectors $\va_i \sim  \mathcal{N}(0,I_n)$ for $i=1,\ldots,m$. For the complex Gaussian case, the signal $\vx \sim  \mathcal{N}(0,I_n)+i  \mathcal{N}(0,I_n)$ and measurement vectors $\va_i  \sim \mathcal{N}(0,I_n/2)+i \mathcal{N}(0,I_n/2)$.  For WF, TWF and TAF, we use the code provided in the original papers with suggested parameters.

\begin{example}{\rm 
In this example, we test the empirical success rate of SAF versus the number of measurements with parameter $\beta=1/2$.  We conduct the experiments for the real and complex Gaussian cases respectively.
We choose $n=128$. The step size $\mu=0.6$ and the maximum number of iterations is $T=2000$.  For the number of measurements, we vary $m$ within the range $[n,8n]$. For each $m$, we run $100$ times trials to calculate the success rate.  Here, we say a trial to have successfully reconstructed the target signal if the relative error satisfies $\mbox{dist}(\vz_{T}-\vx)/\norm{\vx} \le 10^{-5}$.
The results are plotted in Figure \ref{figure:succ}. It can be seen that $4.5n$ real Gaussian phaseless measurement  or $5.5n$ complex Gaussian phaseless measurement are enough for exactly recovery for SAF.

\begin{figure}[H]
\centering
    \subfigure[]{
     \includegraphics[width=0.45\textwidth]{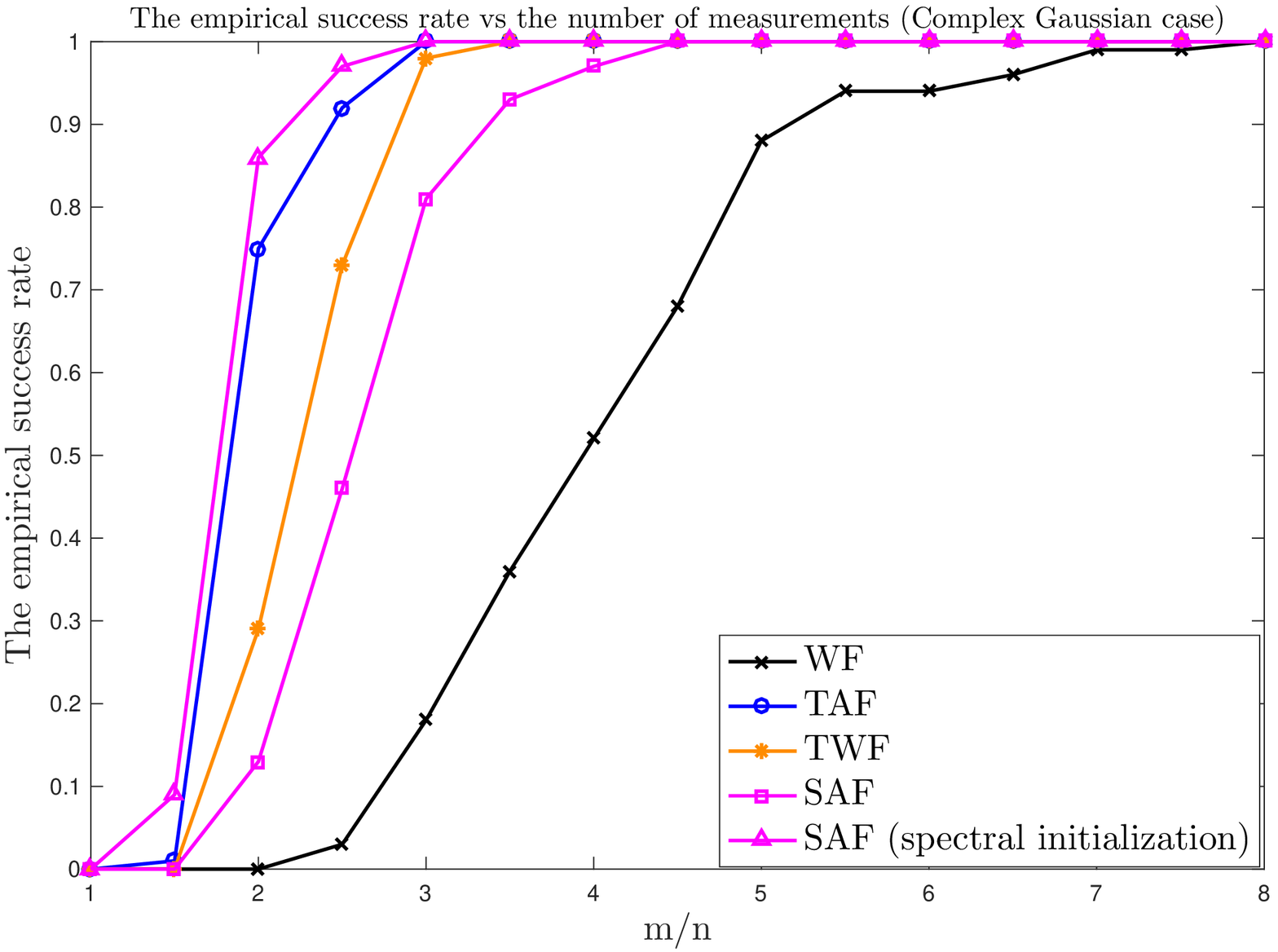}}
\subfigure[]{
     \includegraphics[width=0.45\textwidth]{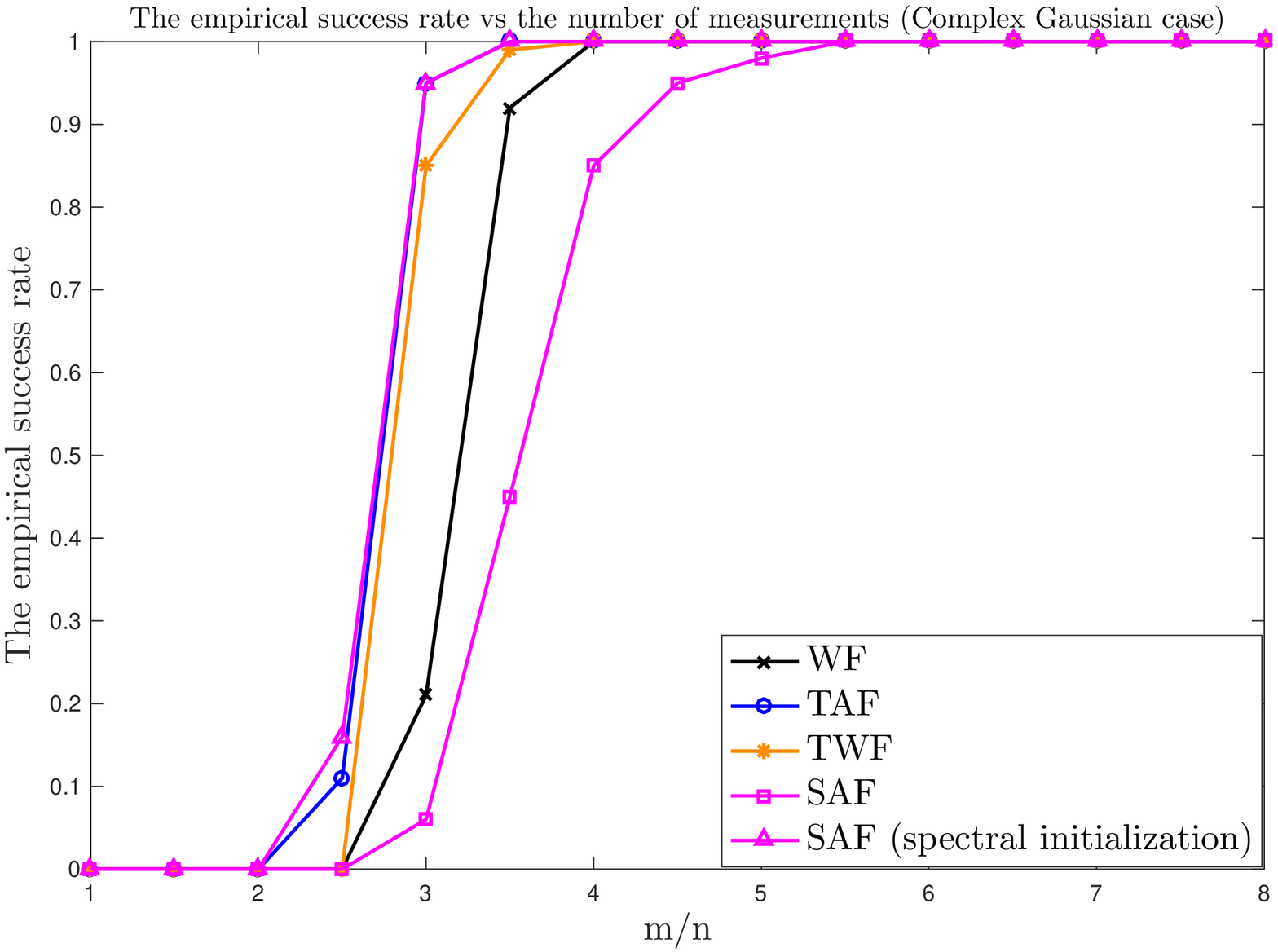}}
\caption{ The empirical success rate for different $m/n$ based on $100$ random trails. (a) Success rate for real Gaussian case, (b) Success rate for complex Gaussian case.}
\label{figure:succ}
\end{figure}

}\end{example}

\begin{example}
{\rm In this example, we compare the convergence rate of SAF with those of WF, TWF, TAF for real Gaussian and complex Gaussian cases. We choose $n=128$ and $m=5n$. The step size $\mu=0.8$ and parameter $\beta=1/2$.  To show the robustness of our SAF, we also consider the noisy data model $y_i=\abs{\nj{\va_i,\vx}}+\eta_i$ where the noise $\eta_i\sim 0.01 \cdot N(0,1)$. The results are presented in Figure \ref{figure:relative_error}. Since our SAF algorithm chooses a random initial guess according to the standard Gaussian distribution instead of adopting a spectral initialization, it sometimes need to escape the saddle points with a small number of  iterations.  Due to its high efficiency to escape the saddle points, it still performs well comparing with state-of-the-art algorithms with spectral initialization. 
\begin{figure}[H]
\centering
\subfigure[]{
     \includegraphics[width=0.45\textwidth]{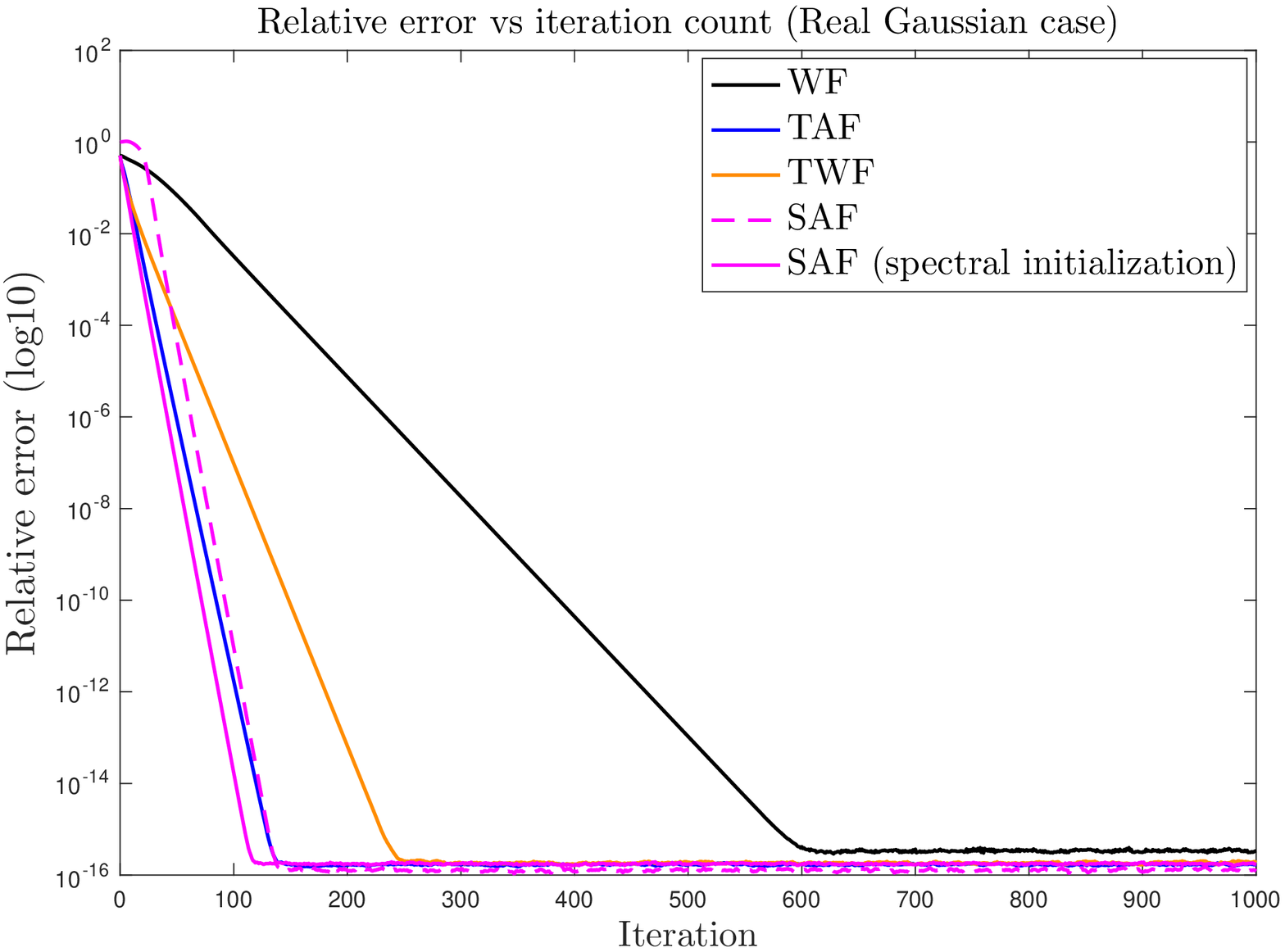}}
\subfigure[]{
     \includegraphics[width=0.45\textwidth]{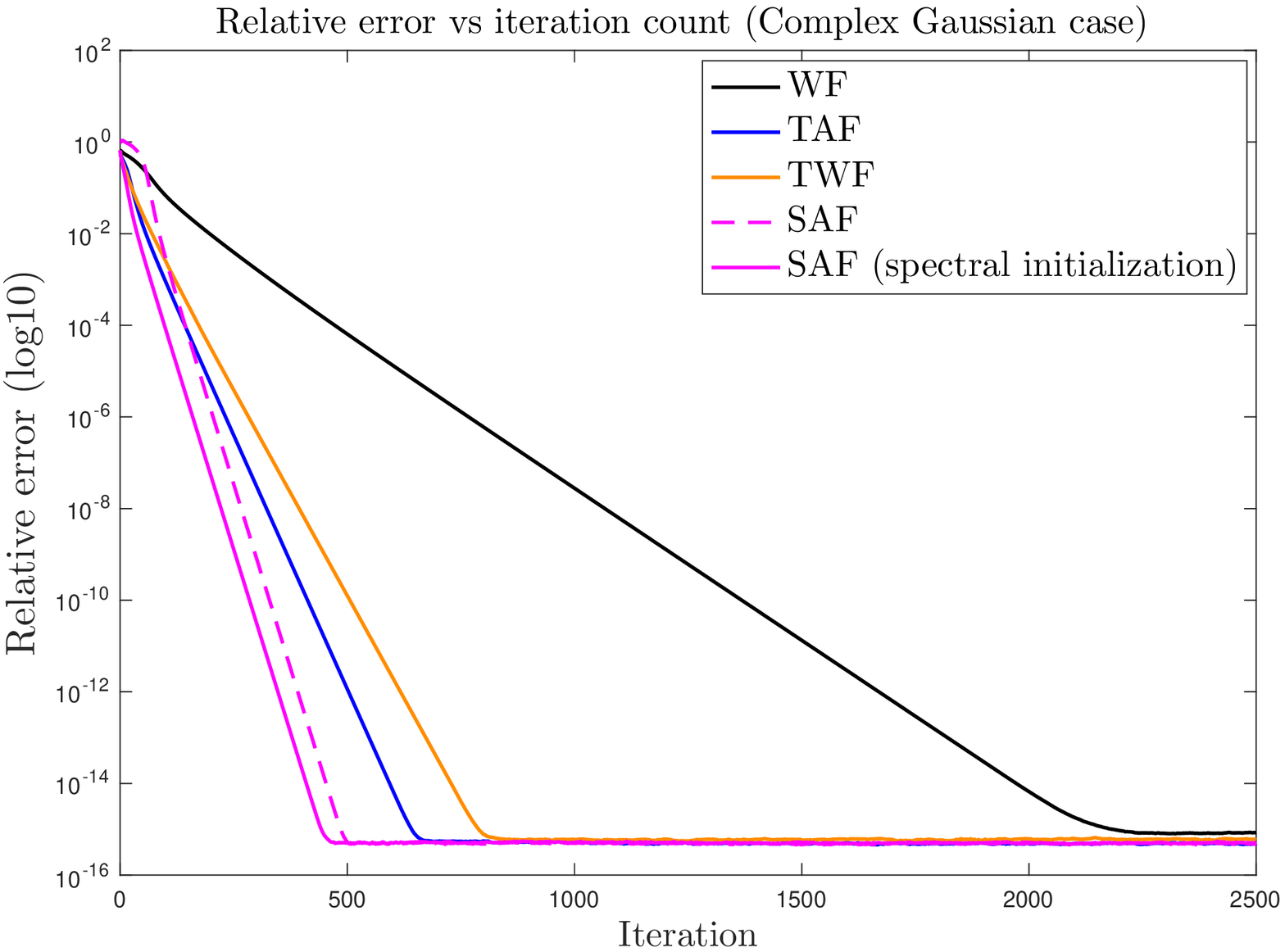}}
     \subfigure[]{
     \includegraphics[width=0.45\textwidth]{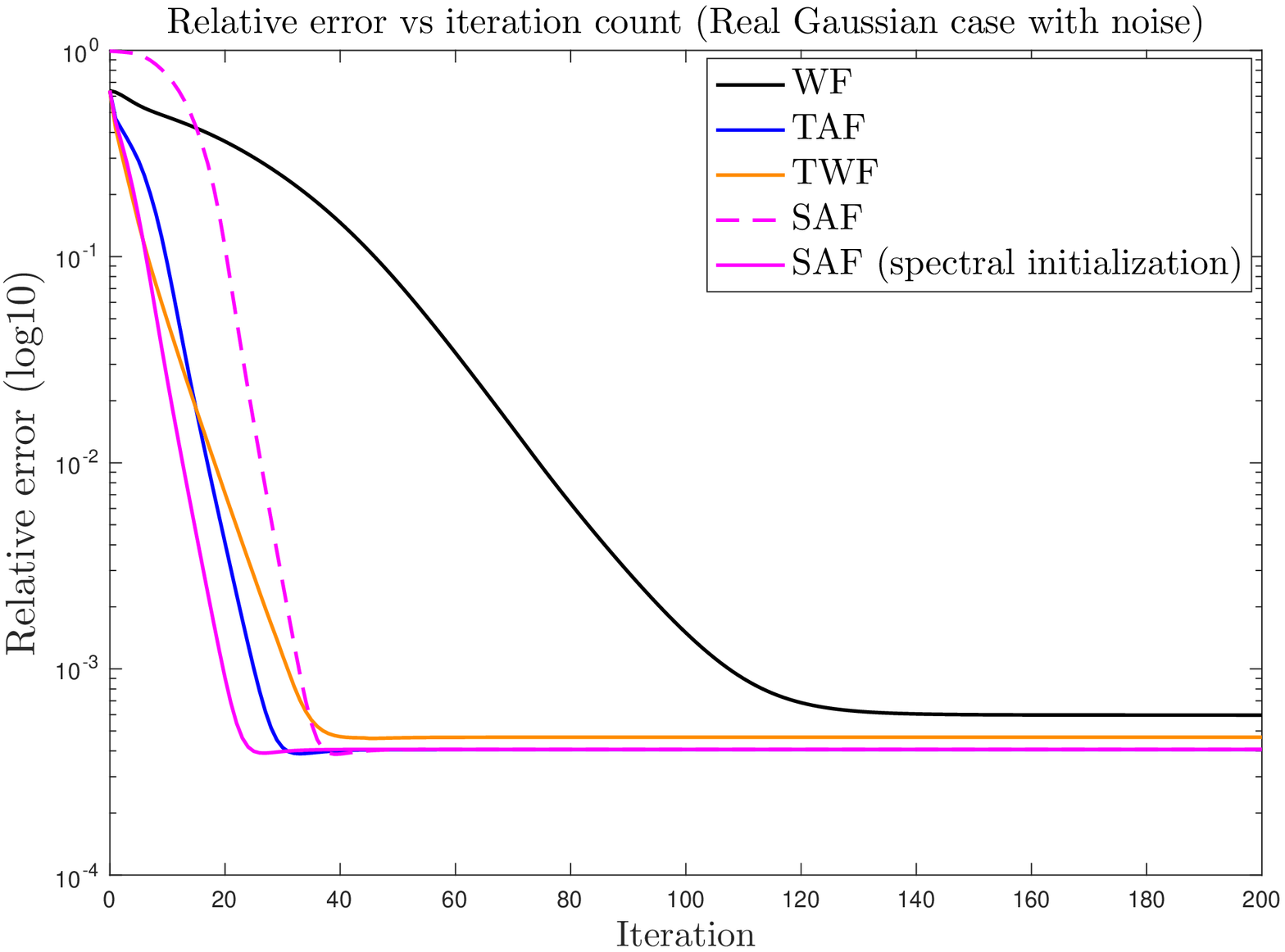}}
\subfigure[]{
     \includegraphics[width=0.45\textwidth]{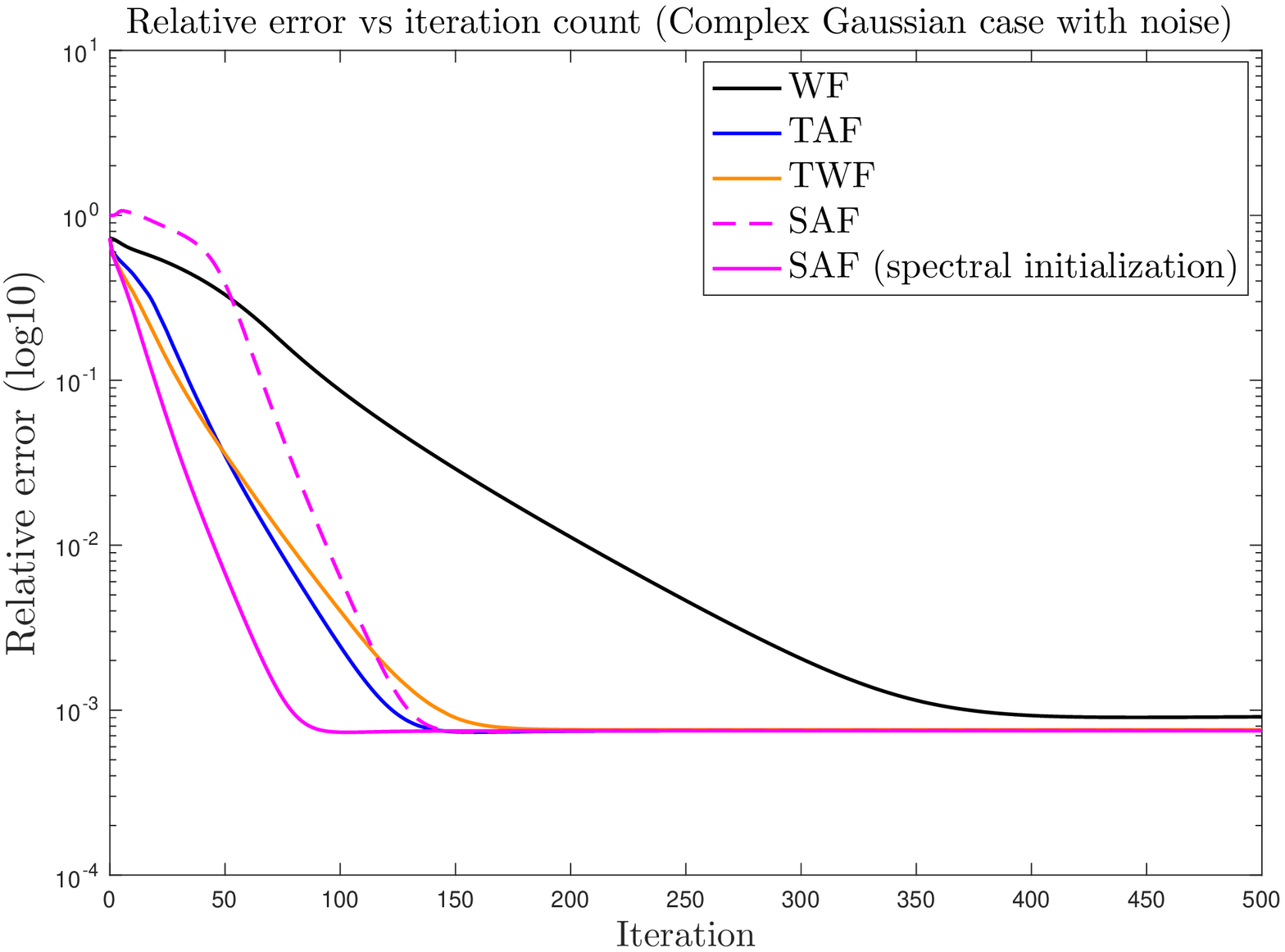}}
\caption{ Relative error versus number of iterations for SAF, WF, TWF, and TAF methods: (a) The noiseless measurements for real Gaussian case; (b) The noiseless measurements for complex Gaussian case; (c) The noisy measurements for real Gaussian case; (d) The noisy measurements for complex Gaussian case.}
\label{figure:relative_error}
\end{figure}
}
\end{example}

\begin{example}{\rm
In this example, we compare the time elapsed and the iteration needed for WF, TWF, TAF and our SAF to achieve the relative error  $10^{-5}$ and $10^{-10}$, respectively.  We choose $n=1000$ with $m=8n$.  The step size $\mu=0.8$. For the parameter $\beta$ in our SAF, we consider the case $\beta=1/2$. We adopt the same spectral initialization method for WF, TWF, TAF and the initial guess is obtained by power method with $50$ iterations.  We run $50$ times trials to calculate the average time elapsed and iteration number for those algorithms. The results are shown in Table \ref{tab:performance_comparison}. The numerical results show that SAF takes around $20$ and $40$ iterations to escape the saddle points for the real and complex Gaussian cases, respectively. Since there is no spectral initialization, the high efficiency of escaping saddle points and low computational complexity, the time elapsed of SAF is less than the other methods significantly. 

}
\end{example}

\begin{table}[tp]
  \centering
  \fontsize{12}{16}\selectfont
  \caption{Time Elapsed and Iteration Number among Algorithms on Gaussian Signals with $n=1000$.}
  \label{tab:performance_comparison}
    \begin{tabular}{|c|c c|cc|cc|cc|}
    \hline
    \multirow{2}{*}{Algorithm}&
    \multicolumn{4}{c|}{Real Gaussian}&\multicolumn{4}{c|}{ Complex Gaussian }\cr\cline{2-9}
    &\multicolumn{2}{c|}{$10^{-5}$ }&\multicolumn{2}{c|}{$10^{-10}$ }& \multicolumn{2}{c|}{$10^{-5}$ } &\multicolumn{2}{c|}{$10^{-10}$ }\cr \hline
    & Iter & Time(s) & Iter & Time(s) & Iter & Time(s) & Iter & Time(s) \cr \hline
   SAF & 44&\bf{0.1556} &68 &\bf{0.2276} &113&\bf{1.3092} & 190 &\bf{2.3596} \cr\hline
   SAF (spectral)&\bf{25}&0.2631& \bf{51} &0.3309 &\bf{67}&1.4528& \bf{151} &2.6122\cr\hline
    WF &125&4.4214& 229 &6.3176 &304&34.6266& 655&86.6993\cr \hline
    TAF &29&0.2744&60&0.3515 &100&1.7704& 211 &2.7852\cr \hline
    TWF&40&0.3181&87&0.4274&112&1.9808& 244&3.7432\cr \hline
    \end{tabular}
\end{table}

\begin{example} {\rm
In this example, we show the performance of SAF with different parameter $\beta$ in the real Gaussian case. We choose $n=128$ and the step size $\mu=0.6$. For SAF with random initialization, we choose $m=4n$; and for SAF with spectral initialization, we choose $m=2.5n$.  We test the parameter $\beta$ within the range from $0.1$ to $1.0$. For each $\beta$, we run $100$ times trials and calculate the success rate, where a trial is successful if the relative error is less than $10^{-5}$.
The results are depicted in Figure \ref{figure:beta}.  From the figures, we can see our SAF has better performance as the parameter $\beta$ increasing.
}
\end{example}
\begin{figure}[H]
\centering
\subfigure[]{
     \includegraphics[width=0.45\textwidth]{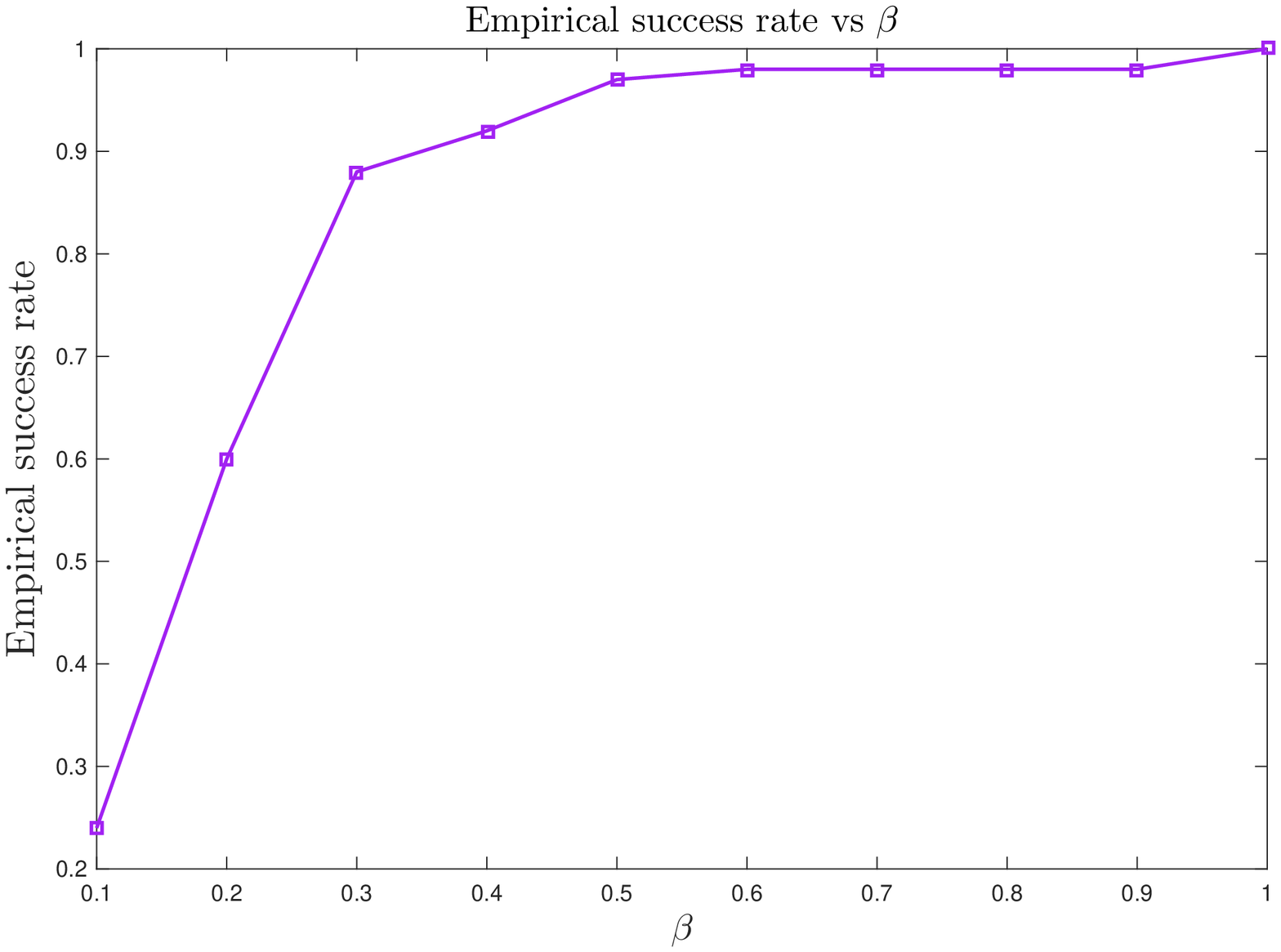}}
\subfigure[]{
     \includegraphics[width=0.45\textwidth]{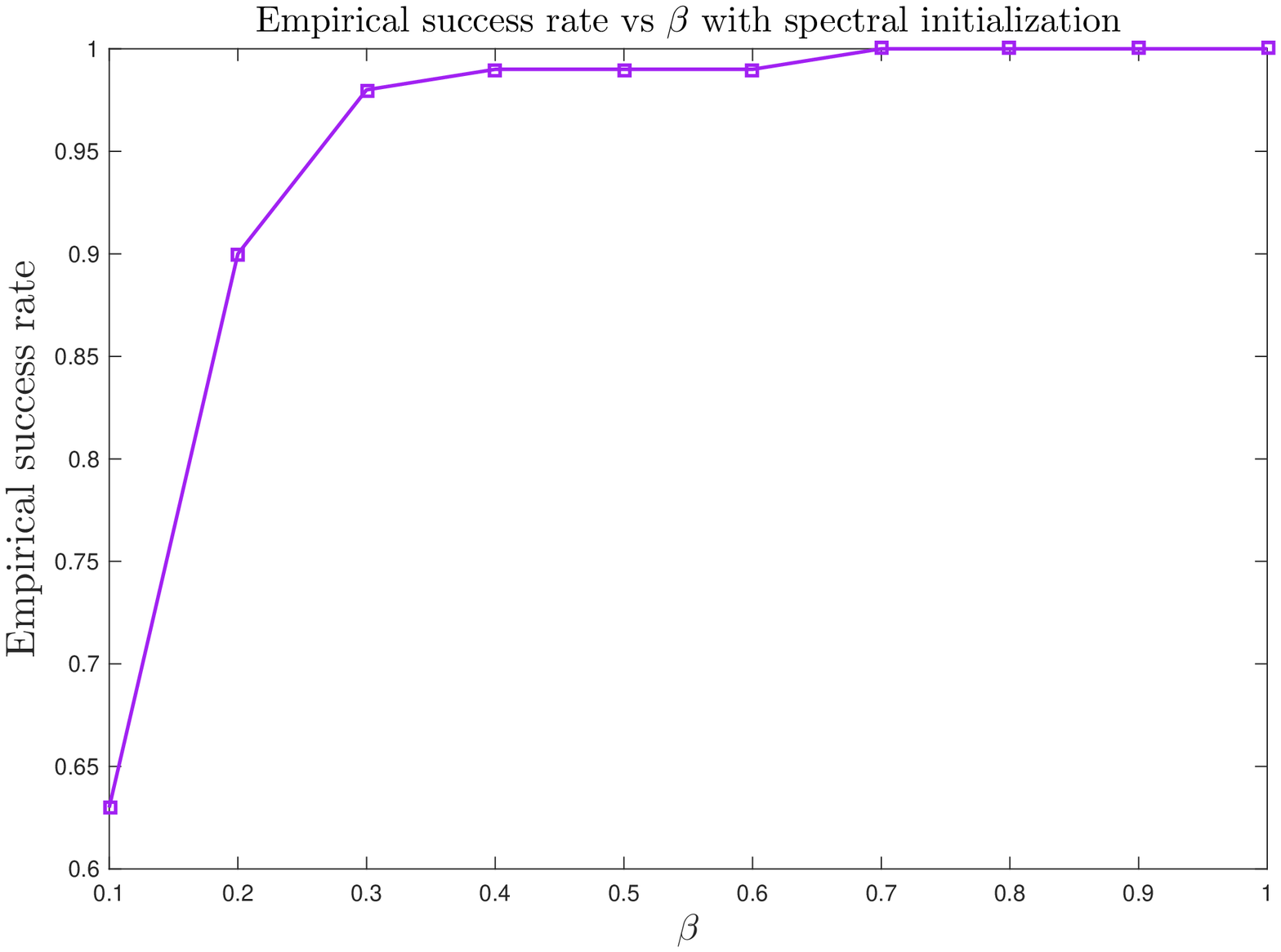}}
\caption{ Relative error versus the parameter $\beta$ for SAF: (a) SAF with random initialization under $m=4n$; (b) SAF with spectral initialization under $m=2.5n$.}
\label{figure:beta}
\end{figure}

\section{Appendix}

\noindent
{\bf Proof of Corollary \ref{coro:union}:}~~
To simplify the presentation, we shall use the  phrase ``with high probability " to mean that the probability is at least $1- C \exp(-c m)$ for some constants $c$, $\delta>0$. Also we shall
tacitly assume that $m\ge C_1 n$ for some sufficiently large constant $C_1$ throughout this
proof. 

For any fixed $\vz_0\in \Omega $, the terms $g(\va_i^\T \vz_0, \va_i^\T \vx)$ are subexponential with subexponential norm $\tau$. By Lemma \ref{lem:concentration}, it holds with
high probability that 
\begin{equation}  \label{eq:concentration11}
      \Bigl|\frac{1}{m} \sum_{i=1}^m  g(\va_i^\T \vz_0, \va_i^\T \vx) - \E[g(\va_1^\T \vz_0, \va_1^\T \vx)] \Bigr| \leq \frac{\varepsilon}{3}.
\end{equation}
To obtain this for all $\vz \in \Omega$ we need a (standard) covering argument. 
 Since $\Omega$ is  compact,  we can construct a $\delta_0$-net $\mathcal{N}$ with
 $\operatorname{Card}(\mathcal N) \le \exp(\alpha_1 n)$ ($\alpha_1$ depends on $\delta_0$)
  such that
  \begin{align*}
  \mathcal N \subset \Omega \subset \bigcup_{\tilde{\vz} \in \mathcal N}
  B(\tilde{\vz}, \delta_0).
  \end{align*}
  Here $\delta_0>0$ is a constant which will be taken sufficiently small. The needed smallness
  will be specified later. 
  By using the above construction,    for any $\vz \in \Omega$, we can find a vector $\vz_0 \in \mathcal{N}$ such that $\norm{\vz-\vz_0}\le \delta_0$. Since $K(\va_i)$ are subgaussian random variables with subgaussian norm $\eta$,  Lemma \ref{lem:concentration} implies that with high probability it holds that
  \[
\frac{1}{m} \sum_{i=1}^m K(\va_i)^2 \le \E K(\va_1)^2+\varepsilon \le \eta^2+\delta_1,
\]
where $\delta_1>0$ will be taken sufficiently small.
On the other hand, we have
\begin{eqnarray} \label{eq:Kai}
       && \left| \frac{1}{m} \sum_{i=1}^m g(\va_i^\T \vz, \va_i^\T \vx)-\frac{1}{m} \sum_{i=1}^m g(\va_i^\T \vz_0, \va_i^\T \vx)\right| \nonumber \\
       & \leq &  \frac{1}{m} \sum_{i=1}^m K(\va_i) \abs{ \va_i^\T (\vz-\vz_0)}   \nonumber \\
      & \leq &  \sqrt{\frac{1}{m} \sum_{i=1}^m K(\va_i)^2  }\cdot \sqrt{\frac{1}{m} \sum_{i=1}^m  \abs{ \va_i^\T (\vz-\vz_0)}^2 } \nonumber\\
       &\le&   \sqrt{ \eta^2+\delta_1} \cdot 2\delta_0,
\end{eqnarray}
where in the last inequality we used the fact that with high probability:
\begin{align*}
\frac 1m \sum_{i=1}^m |\va_i^\T \tilde{\vz} |^2\le 1.01,
\qquad\forall\, \tilde{\vz} \in \mathbb S^{n-1}.
\end{align*}

Furthermore,
\begin{equation} \label{eq:expectgaai}
\abs{\E g(\va_1^\T \vz, \va_i^\T \vx)-\E g(\va_1^\T \vz_0, \va_i^\T \vx)} \le  \E\zkh{ K(\va_1) \abs{ \va_1^\T (\vz-\vz_0)}} \le  \eta \delta_0.
\end{equation}
Choose $\delta_0=\frac{\sqrt{2}\varepsilon}{12\eta}$ and $\delta_1=\eta^2$.  Taking the union bound together with (\ref{eq:concentration11}), (\ref{eq:Kai}) and (\ref{eq:expectgaai}),  we obtain the following: with high probability    it holds that 
\begin{eqnarray*}
 && \Bigl|\frac{1}{m} \sum_{i=1}^m  g(\va_i^\T \vz, \va_i^\T \vx) - \E[g(\va_1^\T \vz, \va_1^\T \vx)] \Bigr| \\
 & \le & \Bigl|\frac{1}{m} \sum_{i=1}^m  g(\va_i^\T \vz, \va_i^\T \vx) - \frac{1}{m} \sum_{i=1}^m  g(\va_i^\T \vz_0, \va_i^\T \vx)\Bigr| + \Bigl|\frac{1}{m} \sum_{i=1}^m  g(\va_i^\T \vz_0, \va_i^\T \vx) - \E[g(\va_1^\T \vz_0, \va_1^\T \vx)] \Bigr| \\
 && \quad +  \Bigl|  \E[g(\va_1^\T \vz, \va_i^\T \vx)] - \E[g(\va_1^\T \vz_0, \va_1^\T \vx)] \Bigr| \\
 &\le &2\delta_0\sqrt{ \eta^2+\delta_1}+ \frac{\varepsilon}{3}+\eta \delta_0 \\
 &\le & \varepsilon
\end{eqnarray*}
 for all  $\vz \in \Omega$. We reminder the reader that we need to take $m\ge C_1 n$ with
 $C_1$ sufficiently large to damp the pre-factor $\alpha_1^n$ in the covering argument.
 \eproof

\noindent
{\bf Proof of Corollary \ref{coro:unidenomn}:}~~
Since $\chi(t)$ is a Lipschitz and compact support function with $\supp(\chi) \subset [0,1]$, it suffices to consider the case where $\abs{\va_i^\T \vz_1} \le \abs{\va_i^\T \vx}$ and $\abs{\va_i^\T \vz_2} \le \abs{\va_i^\T \vx}$ for all $i$. 
Thus we have
\begin{eqnarray*}
&& \Biggl|\frac{1}{m} \sum_{i=1}^m  \frac{\abs{\va_i^\T \vz_1}}{\abs{\va_i^\T \vx}}\chi\Big(\frac{\abs{\va_i^\T \vz_1}}{\abs{\va_i^\T \vx}}\Big) - \frac{\abs{\va_i^\T \vz_2}}{\abs{\va_i^\T \vx}}\chi\Big(\frac{\abs{\va_i^\T \vz_2}}{\abs{\va_i^\T \vx}}\Big) \Biggr| \\
&\le & \frac{1}{m} \sum_{i=1}^m \left| \chi\Big(\frac{\abs{\va_i^\T \vz_1}}{\abs{\va_i^\T \vx}}\Big) - \chi\Big(\frac{\abs{\va_i^\T \vz_2}}{\abs{\va_i^\T \vx}}\Big) \right| +  \frac{1}{m} \sum_{i=1}^m \left| \frac{\abs{\va_i^\T \vz_1}}{\abs{\va_i^\T \vx}} - \frac{\abs{\va_i^\T \vz_2}}{\abs{\va_i^\T \vx}} \right| \\
&\le &  \frac{L+1}{m} \sum_{i=1}^m \frac{\abs{\va_i^\T (\vz_1-\vz_2)}}{\abs{\va_i^\T \vx}}\\
&=&  \frac{L+1}{m} \sum_{i=1}^m \frac{\abs{\va_i^\T (\vz_1-\vz_2)}}{\abs{\va_i^\T \vx}} \1_{\abs{\va_i^\T \vx} \ge \delta }+ \frac{L+1}{m} \sum_{i=1}^m \frac{\abs{\va_i^\T (\vz_1-\vz_2)}}{\abs{\va_i^\T \vx}} \1_{\abs{\va_i^\T \vx} < \delta } \\
&\le & \frac{L+1}{\delta} \cdot  \frac{1}{m} \sum_{i=1}^m \abs{\va_i^\T (\vz_1-\vz_2)} + 2(L+1) \cdot \frac{1}{m} \sum_{i=1}^m  \1_{\abs{\va_i^\T \vx} < \delta } .
\end{eqnarray*}

{For 
 the first term, if $m\ge C\epsilon^{-2} n$ then with probability at least $1-2\exp(-c \epsilon^2 m)$ it holds that
\[
\frac{1}{m} \sum_{i=1}^m \abs{\va_i^\T (\vz_1-\vz_2)}  \le \sqrt{\frac{1}{m} \sum_{i=1}^m \abs{\va_i^\T (\vz_1-\vz_2)}^2}  \le  \norm{\vz_1-\vz_2} +\epsilon.
\]
}

For the second term, note that
\[
\E  \zkh{\1_{\abs{\va_i^\T \vx} < \delta }}= \frac{1}{\sqrt{2\pi}} \int_{-\delta}^\delta e^{-u^2} du \le  \frac{2\delta}{\sqrt{2\pi}}.
\]
Since $\1_{\abs{\va_i^\T \vx} < \delta } $ are bounded random variables, applying  the Hoeffding's inequality gives
\[
\frac{1}{m} \sum_{i=1}^m  \1_{\abs{\va_i^\T \vx} < \delta } \le  \frac{2\delta}{\sqrt{2\pi}}+\epsilon
\]
with  probability at least $1-2\exp(-c\epsilon^2 m)$.

Combining the two estimations gives the conclusion.
\eproof

\vspace{3mm}
\noindent
{\bf Proof of Proposition \ref{lem:dlambda}:}~~
Observe that $U=\sigma V+\tau W$. Then  the condition $|U| \leq \lambda |V|$ is equivalent to 
\begin{equation*} 
     A= \Bigl\{\tau^{-1}(-\lambda -\sigma\, \sgn(V))|V| \leq W \leq \tau^{-1}(\lambda -\sigma \,\sgn(V))|V|\Bigr\}.
\end{equation*}
From the definition, we have
\begin{eqnarray*}
G(\lambda)=\E\zkh{g(U,V)\1_A}&=& \frac{1}{2\pi} \int \int_{|\sigma v+\tau w | \leq \lambda |v|}  g(\sigma v+\tau w,v) e^{-\frac{1}{2}(v^2+w^2)} dwdv \\
&=&  \frac{1}{2\pi} \int_0^\infty \int_{-\xkh{\frac{\lambda}{\tau}+\frac{\sigma}{\tau}}v}^{\xkh{\frac{\lambda}{\tau}-\frac{\sigma}{\tau}}v} g(\sigma v+\tau w,v) e^{-\frac{1}{2}(v^2+w^2)} dwdv \\
&& + \frac{1}{2\pi} \int_{-\infty}^0 \int_{\xkh{\frac{\lambda}{\tau}-\frac{\sigma}{\tau}}v}^{-\xkh{\frac{\lambda}{\tau}+\frac{\sigma}{\tau}}v} g(\sigma v+\tau w,v) e^{-\frac{1}{2}(v^2+w^2)} dwdv.
\end{eqnarray*}
It gives that
\begin{eqnarray*}
G'(\lambda) &=& \frac{1}{2\pi\tau} \int_0^\infty  g(\lambda v,v) v e^{-\frac{1}{2} \mu_-^2v^2} dv+ \frac{1}{2\pi\tau} \int_0^\infty  g(-\lambda v,v) v e^{-\frac{1}{2}\mu_+^2v^2} dv\\
&&  \mathrel{\phantom{=}}- \frac{1}{2\pi\tau}  \int_{-\infty}^0  g(-\lambda v,v) v e^{-\frac{1}{2} \mu_+^2v^2} dv- \frac{1}{2\pi\tau}  \int_{-\infty}^0 g(\lambda v,v) v e^{-\frac{1}{2}\mu_-^2v^2} dv\\
&=& \frac{1}{ 2\pi\tau}\int_0^\infty\left(g(-\lambda v,v)+g(\lambda v,-v)\right)ve^{-\frac{1}{2}\mu_+^2v^2}dv \\
    && \mathrel{\phantom{=}}  +  \frac{1}{ 2\pi\tau}\int_0^\infty\left(g(\lambda v,v)+g(-\lambda v,-v)\right)ve^{-\frac{1}{2}\mu_-^2v^2}dv. 
\end{eqnarray*}
\eproof

\vspace{3mm}
\noindent
{\bf Proof of Lemma \ref{lem:EUV-VW}:}~~
For the expectation $ \E[|UV|] $, let $\sigma=\cos\alpha $ for some $\alpha\in [0,2\pi)$. Then
we have
\begin{eqnarray*}
 \E[|UV|]  &=& \frac{1}{2\pi} \int_{-\infty}^\infty \int_{-\infty}^{\infty }  \abs{v (\sigma v+\tau w)} \cdot e^{-\frac{1}{2}(v^2+w^2)} dwdv \\
 &=& \frac{1}{2\pi} \int_{0}^{2\pi}  \int_{0}^\infty   r^3 \abs{\sin \theta} \cdot \abs{\sigma \sin\theta +\tau \cos \theta} \cdot e^{-\frac{1}{2}r^2} dr d\theta \\
 &=& \frac{1}{\pi}\int_{0}^{2\pi}  \abs{\sin\theta \sin (\theta+\alpha)}d\theta\\
  &=& \frac{2}{\pi}\xkh{\sin \alpha + (\frac{\pi}{2}-\alpha)\cos \alpha} \\
   &=& \frac{2}{\pi}\xkh{ \tau+\sigma\arctan\frac{\sigma}{\tau}}.
\end{eqnarray*}

Finally,  
\begin{eqnarray*}
 \E[\sgn(UV)V^2)] &=& \frac{1}{2\pi} \int_{-\infty}^\infty \int_{-\infty}^{\infty }  \sgn(\sigma v^2+\tau vw) \cdot v^2 \cdot e^{-\frac{1}{2}(v^2+w^2)} dwdv \\
 &=& \frac{1}{\pi}\int_{0}^{2\pi}  \sgn\xkh{\sin\theta \sin (\theta+\alpha)} \sin^2\theta d\theta\\
  &=& \frac{2}{\pi}\int_{0}^{\pi-\alpha}   \sin^2\theta d\theta -\frac{2}{\pi}\int_{\pi-\alpha}^{\pi}   \sin^2\theta d\theta \\
  &=& \frac{2}{\pi}\xkh{\sin (2\alpha) + \frac{\pi}{2}-\alpha} \\
   &=& \frac{2}{\pi}\xkh{ \tau\sigma+\arctan\frac{\sigma}{\tau}}.
\end{eqnarray*}

\eproof

\vspace{3mm}
\noindent
{\bf Proof of bounds and Lipschitz property of $\Psi_u$:}~~
The bounds and Lipschitz property of $\Psi_u$ given in (\ref{eq:Psi_bound}) and (\ref{eq:Psi_Lip}) are easy to check and we only prove the bound
\begin{equation}  \label{eq:psilowbound}
 \Psi_u(u,v)u \geq u^2- |uv|.
\end{equation}
Recall the expression (\ref{eq:partial_Psi}) of $ \Psi_u(u,v)$. We have
\begin{equation*}  
    \frac{\Psi_u(u,v)u}{v^2}
        = \left \{ \begin{array}{ll}
           \xkh{ \frac{u}{v}}^2-\abs{ \frac{u}{v}}, &  \abs{u} > \beta |v| ; \\
    \frac{1}{2\beta^2}\cdot \xkh{ \frac{u}{v}}^4 + (\frac{1}{2}-\frac{1}{\beta})\xkh{ \frac{u}{v}}^2, &  \abs{u} \leq \beta |v|. \end{array} \right.
\end{equation*}
Thus, to prove (\ref{eq:psilowbound}) it suffices to show
\[
 \frac{1}{2\beta^2} t^4 +(\frac{1}{2}-\frac{1}{\beta}) t^2\ge t^2-\abs{t} 
\]
for all $\abs{t}\le \beta$. We only need to consider $0\le t \le \beta$. Let
\[
h(t)=t^3-(\beta^2+2\beta) t + 2\beta^2.
\]
Then $h(t)$ is a decreasing function for $0\le t \le \beta$. Note that $h(\beta)=0$. It gives $h(t)\ge 0$ for all $0\le t \le \beta$. We arrive at the conclusion.
\eproof

\section*{Proof of a more general inequality}
Denote $\tau= \sqrt{1-\sigma^2}$ and $0\le \sigma \le 1$. We shall show
that
\begin{align*}
f_0(\tau) = \frac 1 {\tau^2} ( \frac {\pi}2
-\tau - \frac 1 {\sigma} \arctan(\sigma/\tau) )
\end{align*}
is monotonically increasing for $0<\tau<1$.  

\underline{First transformation}.  Write
\begin{align*}
f_0(\tau) &= \tau^{-2} ( \frac {\pi}2 -\tau - \int_{\tau}^{\infty}
\frac 1 {s^2 +1-\tau^2} ds ) \notag \\
&=\tau^{-2} (  \int_0^{\infty} (\frac 1 {s^2+1} -\frac 1 {s^2+1-\tau^2})ds
+\int_0^{\tau}(\frac 1 {s^2+1-\tau^2} -1) ds ) \notag \\
&=-\int_0^{\infty} \frac 1 {(1+s^2)(1+s^2-\tau^2)} ds
+\tau \int_0^1 \frac{1-s^2} {1-(1-s^2) \tau^2} ds.
\end{align*}
Denote $t=\tau^2$ and 
\begin{align*}
\tilde f_0(t) = -\int_0^{\infty} \frac 1 {(1+s^2)(1+s^2-t)} ds
+\sqrt t\int_0^1 \frac{1-s^2} {1-(1-s^2) t} ds.
\end{align*}
Note that
\begin{align*}
&f_0({\tau})=\tilde f_0({\tau}^2), \\
&f_0^{\prime}(\tau) = 2\tau \tilde f_0^{\prime}(\tau^2).
\end{align*}
We have
\begin{align*}
\tilde f_0^{\prime}(t)
&= -\int_0^{\infty} \frac 1 {(1+s^2)(1+s^2-t)^2} ds
+\sqrt t\int_0^1 \frac{(1-s^2)^2} { (1-(1-s^2) t)^2} ds \notag \\
&\qquad +\frac 1 2 t^{-\frac 12} \int_0^1 \frac{1-s^2} {1-(1-s^2) t} ds.
\end{align*}

\underline{Second transformation}. 
Our second transformation is to
 set $\tau=\cos \theta$, $\sigma= \sin \theta$. Then 
\begin{align*}
-f_0(\cos \theta)=f_1(\theta) = \frac { \frac {\theta}{\sin \theta} + \cos \theta -
\frac {\pi}2 } {\cos^2 \theta}
\Bigl( \text{for Remark \ref{rem_tmp1} }\: = \frac {1+ (\cos \theta - \frac {\pi}2) \frac{\sin \theta}{\theta}} 
{\cos^2\theta} \cdot \frac {\theta}{\sin \theta} \Bigr).
\end{align*}
is increasing on $[0, \frac {\pi}2]$.  Note that $\frac{\theta}{\sin \theta}$ is
monotonically increasing.

\medskip
\medskip
\begin{remark} \label{rem_tmp1}
As another variant, one can  consider
\begin{align*}
f(t,\theta)= \frac { t (1-\frac{\pi}2 \frac{\sin \theta}{\theta})
+ \frac{\cos \theta \sin \theta} {\theta}}
{\cos^2 \theta}.
\end{align*}
For $0\le  t\le 1$, $f(t, \theta)$ is an increasing function of $\theta$.
For $t>1$, this does not hold. Note that $\frac {\theta}{\sin \theta}$ is monotonically
increasing on $[0, \frac {\pi}2]$. The monotonicity of $f(1,\theta)$ would yield the
monotonicity of $f_1$. 
\end{remark}

\medskip
\medskip
Since
\begin{align*}
f_1(\theta) = \frac { \frac {\theta}{\sin \theta} + \cos \theta -
\frac {\pi}2 } {\cos^2 \theta}=-f_0(\cos \theta),
\end{align*}
we have
\begin{align*}
f_1^{\prime}(\theta) = \sin \theta f_0^{\prime}(\cos \theta).
\end{align*}
The derivative of $f_1(x)$ is
\begin{align*}
f_1^{\prime}(x)= \sec x ( -x \csc^2 x +2x \sec^2 x +\csc x \sec x +\tan x -\pi \sec x \tan x).
\end{align*}
The task is to show $f_1^{\prime}(x)\ge 0$ for $x\in [0, \frac {\pi}2]$, i.e.
\begin{align*}
\frac 1 {\cos x} \bigl(\frac{2-3\cos^2 x} {\sin^2 x \cos^2 x}
x + \frac 1 {\sin x \cos x} +\frac{\sin x}{\cos x} - \pi \frac{\sin x }{\cos^2 x}\bigr)\ge 0.
\end{align*}
Denote $\sqrt t= {\cos x}$ so that $\sin x = \sqrt{1-t}$. Then we only need to show
for all $t\in [0,1]$:
\begin{align*}
t^{-\frac 32} (1-t)^{-\frac 12}\biggl( (2-3t)\frac{\arcsin(\sqrt{1-t})}{\sqrt{1-t}}
+(2-t)\sqrt t -\pi(1-t) \biggr)\ge 0.
\end{align*}
Note that
\begin{align*}
f_1^{\prime} \Bigr|_{\cos \theta = \sqrt t} = \sqrt{1-t} f_0^{\prime}(\sqrt t)
= \sqrt{1-t} \cdot  2 \sqrt t\tilde f_0^{\prime}(t).
\end{align*}

Our main idea is to use $f_1^{\prime}$  and $\tilde f_0^{\prime}$ in different regimes.

Case 1: $\frac 23 \le t \le 1$.  Note that $\frac{\arcsin x}x \ge 1$ for $x\in [0,1]$. Easy to check
that in this regime
\begin{align*}
2-3t +(2-t)\sqrt t -\pi (1-t)\ge 0.
\end{align*}
The above is equivalent to checking for $\frac 23 \le t \le 1$:
\begin{align*}
2+\sqrt t + \frac {\sqrt t}{1+\sqrt t} - \pi \ge 0.
\end{align*}
which is obvious thanks to monotonicity.

Case 2: $0<t \le \frac 14 $.  In this case we work with $\tilde f_0^{\prime}(t)$. 
Note that for $0<t \le \frac 14$, 
\begin{align*}
\tilde f_0^{\prime}(t)
&> - \int_0^{\infty} \frac 1 {(1+s^2)(1+s^2-\frac 14)^2} ds
+\sqrt t\int_0^1 {(1-s^2)^2} { (1+(1-s^2) t)^2} ds \notag \\
&\qquad +\frac 1 2 t^{-\frac 12} \int_0^1 {(1-s^2)} {(1+(1-s^2) t)} ds.
\end{align*}
We have
\begin{align*} 
 \int_0^{\infty} \frac 1 {(1+s^2)(1+s^2-\frac 14)^2} ds =
 \frac 9 {16} (8-3\sqrt 6) \pi \approx 0.94875.
 \end{align*}
Obviously then
\begin{align*}
\tilde f_0^{\prime}(t) &>-0.94876 t^{\frac 12} + \frac 13 + \frac {4}5 t +\frac {32}{35}t^2 +
\frac{128}{315} t^3 \notag \\
& > -0.94876 t^{\frac 12} + \frac 13 + \frac {4} 5 t >0, \qquad \forall\, 0<t \le \frac 14.
\end{align*}

Case 2a (slightly better): $0<t \le \frac 13 $.  In this case we still work with $\tilde f_0^{\prime}(t)$. 
Note that for $0<t \le \frac 13$, 
\begin{align*}
\tilde f_0^{\prime}(t)
&> - \int_0^{\infty} \frac 1 {(1+s^2)(1+s^2-\frac 13)^2} ds
+\sqrt t\int_0^1 {(1-s^2)^2} { (1+(1-s^2) t)^2} ds \notag \\
&\qquad +\frac 1 2 t^{-\frac 12} \int_0^1 {(1-s^2)} {(1+(1-s^2) t 
+((1-s^2)t)^2)} ds.
\end{align*}
We have
\begin{align*} 
 \int_0^{\infty} \frac 1 {(1+s^2)(1+s^2-\frac 13)^2} ds =
 \frac 89 (9-5\sqrt 3) \pi \approx 1.15135.
 \end{align*}
Obviously then
\begin{align*}
\tilde f_0^{\prime}(t) &>-1.15136 t^{\frac 12} + \frac 13 + \frac {4}5 t +\frac {8}{7}t^2 
 \notag \\
& > -1.15136 t^{\frac 12} + \frac 13 + \frac {4} 5 t 
+\frac{8} {7} t^2>0, \qquad \forall\, 0<t \le \frac 13.
\end{align*}

We note that one can deal with $t^{\frac 12}$ for $ \frac 14 \le t \le \frac 13$ in the
following way:
\begin{align*}
t^{\frac 12} > \frac 12 + (t-\frac 14) -(t-\frac 14)^2, \quad\forall\, \frac 14 \le t\le \frac 13.
\end{align*}
Then
\begin{align*}
\tilde f_0^{\prime}(t)
&>-1.15136 (\frac 12 + (t-\frac 14) -(t-\frac 14)^2)+ \frac 13 + \frac {4} 5 t 
+ t^2 \notag \\
&> 0.117-0.93 t+2t^2>0, \quad \forall\, \frac 14 \le t \le \frac 13.
\end{align*}

Case 3: $\frac 13 <t<\frac 23$.  Now we shall work with the expression
\begin{align*}
 (2-3t)\frac{\arcsin(\sqrt{1-t})}{\sqrt{1-t}}
+(2-t)\sqrt t -\pi(1-t) \ge 0.
\end{align*}
Make a change of variable $t\to 1-s$ and note that the regime is invariant. We then
need to show for $\frac 13 <s< \frac 23$,
\begin{align*}
h(s)= (3s-1) \frac{\arcsin(\sqrt s)}{\sqrt s} +(1+s)\sqrt{1-s} - \pi s\ge 0.
\end{align*}
It is easy to check that (note that $\arcsin$ has positive-coefficient power series expansion!)
\begin{align*}
\frac{\arcsin(\sqrt s)}{\sqrt s} 
\ge 1+ \frac s 6 + \frac {3}{40} s^2.
\end{align*}
Thus we need to show for $\frac 13 \le s \le  \frac 23$:
\begin{align*}
h(s)= (3s-1) (  1+ \frac s 6 + \frac {3}{40} s^2)+(1+s)\sqrt{1-s} - \pi s\ge 0.
\end{align*}

To this end we rewrite 
\begin{align*}
h(s) 
&= (3s-1) (  1+ \frac s 6 + \frac {3}{40} s^2) - \pi s +(1+s)+
(1+s)(\sqrt{1-s} - 1) \notag \\
&=\frac 1 {120} s(460-120\pi +51 s+27 s^2)  - \frac {s(1+s)} {1+\sqrt{1-s}} \notag \\
&= s( A(s) - \frac{1+s}{1+\sqrt{1-s}}),
\end{align*}
where $A(s)= \frac 1 {120} (460-120\pi +51 s+27 s^2)$.
It is not difficult to check that $0<A(s)<1+s$ (this holds for $0\le s \le 1$) .  Then to show
\begin{align*}
\sqrt{1-s}> \frac{1+s}{A(s)}-1,
\end{align*}
we can square on both sides and then multiply both sides by $A(s)^2$. We then need to check
the inequality
\begin{align*}
g_1(s)=A(s)^2 (1-s) - (1+s-A(s))^2>0.
\end{align*}
After a tedious computation, we obtain
\begin{align*}
g_1(s)&=-\frac{81 s^5}{1600}-\frac{153 s^4}{800}-\frac{2329 s^3}{1600}+\frac{9 \pi  s^3}{20}-\frac{71 s^2}{24}+\frac{17 \pi  s^2}{20}-\pi ^2 s-\frac{368 s}{45}+\frac{17 \pi  s}{3}+\frac{20}{3}-2 \pi \notag \\
&\approx -0.050625 s^5-0.19125 s^4-0.0419083 s^3-0.28798 s^2-0.245024 s+0.383481.
\end{align*}
Clearly $g_1$ is monotonically decreasing and it suffices for us to show $g_1(\frac 23)>0$.
Indeed
\begin{align*}
g_1(\frac 23)\approx 0.035>0.
\end{align*}

\end{document}